\documentclass[a4paper,11pt]{article}
\pdfoutput=1 % if your are submitting a pdflatex (i.e. if you have
\usepackage{jinstpub} % for details on the use of the package, please
\usepackage{subfigure}
\usepackage{lineno}
\title{Simulation of neutron background for a dark matter search experiment at JUSL}

\author[a,1]{S. Banik\note{Corresponding author}}
\author[a]{V. K. S. Kashyap}
\author[b]{S. Ghosh}
\author[b]{S. Dutta}
\author[a]{B. Mohanty}
\author[a]{Meghna K. K.}
\author[b]{\\P. Bhattacharjee}
\author[b]{S. Saha}

\affiliation[a]{School of Physical Sciences, National Institute of Science Education and Research, HBNI,  Jatni-752050, India}
\affiliation[b]{Saha Institute of Nuclear Physics, HBNI, 1/AF Bidhannagar, Kolkata-700064, India}
\emailAdd{samir.banik@niser.ac.in}

\abstract{Dark matter search experiments demand low to ultralow radiation background to operate. It is very important to understand the nature of the radiation background including knowledge about the sources contributing to it. Sometimes, evaluation of the background becomes very specific to the site chosen for the experiment, and also to the experimental configuration. A dark matter search experiment is proposed to be set up at the Jaduguda Underground Science Laboratory (JUSL) in India. The laboratory will be located inside an existing mine with 555 m of vertical rock overburden. Neutrons produced from $(\alpha,n)$ reactions, spontaneous fission of natural radioactive impurities in the rocks, and also from cosmic muon induced reactions are considered as the main background which can affect the sensitivity and outcome of the experiment. In this work, simulations based on GEANT4 are done to understand both the radiogenic neutron background caused by natural radioactivity of the surrounding rock and the cosmogenic neutron background due to interaction of the deeply penetrating cosmic muons with the rock material. The muon flux in the cavern is obtained to be $4.49(\pm0.25)\times10^{-7}$ cm$^{-2}\thinspace$s$^{-1}$ and the fluxes of radiogenic and cosmogenic neutrons above an energy threshold of 1 MeV in the cavern are obtained to be $5.75(\pm0.58)\times10^{-6}\thinspace$cm$^{-2}\thinspace$s$^{-1}$ and $7.25(\pm0.65)\times 10^{-9}\thinspace$cm$^{-2}\thinspace$ s$^{-1}$ respectively. The values obtained are comparable with estimates and measurements done for DAMA, WIPP and dark matter experiments at Boulby mine. The effectiveness of different shielding materials are also investigated to obtain the best possible neutron background reduction for a dark matter search experiment at JUSL. We also estimate the sensitivity of a CsI 
based detector for Weakly Interacting Massive Particle (WIMP) dark matter search at JUSL considering the estimated neutron background.}


\begin{document}
\maketitle
\section{\label{sec:intro}Introduction}
Observations of the cosmic microwave background anisotropies indicate that our universe contains only about 5\% of normal or luminous matter, about 27\% dark matter (DM), and the rest as dark energy which accounts for the accelerated expansion of the universe \cite{planck}. DM, as the name implies, has the characteristic feature that it rarely interacts with normal matter, if at all, to produce measurable signal in a detector. The DM particles can interact with (scatter off) the electrons or nuclei in the detector medium. The resulting recoiling nuclei or electrons would dump their energy in the detector material producing scintillation, ionization and also heat pulses (such as phonons in solid) \cite{LBaudis,schumann} depending on the chosen detector material.

It is well known that there will also be electron recoils generated by gamma rays, beta particles and other ambient radiation quanta which will produce undesirable background signals. Neutron is an important background that affects the outcome of highly sensitive experiments such as direct search for DM\cite{LBaudis}, neutrinoless double beta decay (NDBD) \cite{CAlduino}, neutrino oscillation experiments \cite{FPAn} etc. Neutrons undergo elastic and inelastic scattering with nuclei of the active medium, resulting in nuclear recoil. The same kind of nuclear recoil may result from the interaction of DM particles. Therefore, neutron background limits the sensitivity of direct dark matter search experiments.

Dark matter search at INO (DINO) is a proposed experiment in India on direct search for dark matter, to be set up at a future facility of the India-based Neutrino Observatory (INO) \cite{Whitepaper}. The experiment will be designed to look for the signature of DM candidates through observation of extremely tiny amount of nuclear recoil or electron recoil events in a suitable crystalline detector medium. Inorganic scintillators, such as Cesium Iodide (CsI) and Gadolinium Galium Aluminium Garnet (${\rm Gd}_3{\rm Ga}_3{\rm Al}_2{\rm O}_{12}$) are possible detector materials to explore for the DINO experiment. Both the scintillators manifest good light yield of around 50 photons$\thinspace$keV$^{-1}$ \cite{csi,ggag}. The first phase of the experiment involves exploration and evaluation of the background effects caused by ambient radiation and other sources at the specific site. To begin with, a small underground laboratory (approximately 5 m $\times$ 5 m $\times$ 2.2 m), named as Jaduguda Underground Science Laboratory (JUSL),  is already established at depth of 555 m ($\sim$1600~mwe vertical rock overburden) in an existing mine of Uranium Corporation of India Limited (UCIL), located at Jaduguda in India for exploring the feasibility of the experiment. The main reason for setting up such experiments at deeper underground sites is to reduce the direct effects of cosmic secondary particles. However, most abundant and deeply penetrating component of the cosmic rays being muons, they can still reach the experimental cavern located underground producing neutrons and hadronic showers. Neutrons are generated by spallation reactions of the muons with the underground rock and shielding materials. Neutrons are also generated by hadronic showers. These neutrons are known as cosmogenic neutrons. While the energy spectrum and flux of these neutrons will depend mostly on the depth of the site and the rock composition, the other sources of neutrons, known as the radiogenic neutrons, caused by the ($\alpha,n$) reactions and spontaneous fission of naturally occurring $^{238}$U, $^{232}$Th and their daughters from the decay chains, will depend on both the natural radioactivity and the elemental composition of the surrounding rocks. Quantitative estimates for the two sources of neutron background are therefore, highly dependent on the site, specifically on the abundance of U/Th and the rock composition. Although the relative flux due to cosmogenic neutrons are a few orders of magnitude less than that of the radiogenic neutrons, cosmogenic neutrons are more energetic and therefore, penetrate deeper inside the passive shielding materials, such as lead, copper and plastic, used in dark matter experiments. It is important to compare these backgrounds for different passive shielding configurations to arrive at the best possible configuration for their effective reduction.

The present study attempts to estimate the neutron background at the JUSL site using the GEANT4 toolkit \cite{geant}, with the sources of radiogenic neutrons generated from standard compilations \cite{Mei} and source of cosmogenic neutrons generated from muons using Gaisser parametrization \cite{gaisser}. Propagation of these neutrons inside the shielding materials are also studied to conclude about the best possible shielding configurations to setup a DM search experiment at the site. In section \ref{sec:Sources}, the various sources of neutrons in underground laboratories are discussed. In section \ref{sec:estgeant}, the estimation of radiogenic, cosmogenic and total neutron flux at the site is explained in detail. In section \ref{sec:shield}, various shielding combinations to reduce the neutron flux for a typical direct dark matter search setup assuming a CsI detector are discussed. The estimated sensitivity of a CsI detector based experiment at JUSL for direct dark matter search is presented in section \ref{sec:sens}. We then conclude with the consolidation of results and how they compare with other experimental locations around the world.

\section{Sources of neutron in underground laboratories\label{sec:Sources}}

Neutrons in underground laboratories are produced in reactions either by natural radioactivity or by cosmic rays muons.
\paragraph{Radiogenic neutrons:}
The radiogenic neutrons are mainly produced in $(\alpha,n)$ reactions caused by the $\alpha$ particles from the U/Th traces present in surrounding rocks and the detector materials. The neutrons produced from spontaneous fission of U or Th and neutron induced fission also contribute to radiogenic neutron flux. The neutron induced fission, being a secondary process, will have negligible contribution to the radiogenic neutron flux. The expected energy spectra of the radiogenic neutrons extend up to about 14 MeV. The radiogenic neutrons from the rock due to ($\alpha, n$) reactions and due to spontaneous fission of $^{238}$U are considered in the present work.
\paragraph{Cosmogenic neutrons:}
Cosmogenic neutrons are produced in cosmic ray muon induced interactions inside rock or shielding materials. Muons produce neutrons via following interactions:
\begin{enumerate}
 \item interaction with nuclei producing nuclear disintegration.
 \item muon capture by nucleus followed by neutron emission.
 \item neutron production by hadrons from muon generated showers.
 \item neutron production by gammas from muon generated electromagnetic showers.
\end{enumerate}
Cosmogenic neutrons have harder energy spectrum with energies upto few GeVs. These neutrons can reach detector after propagating large distance, 
away from the parent muon track. Because of their energy, it is very hard to shield them. 
Some of these neutrons can be vetoed if the associated muons give hit in the muon veto. The neutrons which are produced inside rock cannot be vetoed. 
\section{Estimation of neutron flux using GEANT4\label{sec:estgeant}}
Simulations have been performed using GEANT4 \cite{geant} version 10.02.
The reference physics list ``\texttt{Shielding}'', which is recommended for shielding applications at high energies, is used for describing the physics processes \cite{shield}. The high energy part of this physics list is taken from \texttt{FTFP\_BERT} physics list. It uses high precision neutron model. It was originally developed for neutron penetration studies and ion - ion collision. It has been used for simulation of high energy calorimetry and underground or low background experiments. Secondary particle production threshold cuts are set to 0.7 mm for gammas and  $e^+/e^-$.
\subsection{Simulation of radiogenic neutron flux\label{sec:alpha}}
For the calculation of neutron yield from $(\alpha,n)$ reactions, we have used the composition of rock surrounding JUSL obtained by wet-chemical, radiometric and Inductively Coupled Plasma - Optical Emission Spectrometry (ICP-OES) analyses \cite{JR} (given in Table~\ref{tab:table1}). The rock sample was collected by core drilling at 555 m depth. ICP-OES and wet-chemical analyses give the elemental composition in the rock whereas the radiometric analysis only gives an estimate of the K, U and Th content of the rock. Due to the variation among rock samples and the results obtained from different methods, the uncertainty in the elemental composition is around 10-20\%. The rock, having a density $\sim2.89\pm0.09$ g$\thinspace$cm$^{-3}$, contains 8 ppm of U and 16 ppm of Th \cite{JR}.
\begin{table}[h]
  \centering
  \caption{\label{tab:table1}The composition of Jaduguda rock as obtained by wet-chemical, radiometric and ICP-OES analyses \cite{JR}.}
\begin{tabular}{cc|cc}
\hline 
Element & Conc (\%) & Element & Conc (\%)\\ \hline
U & 0.0008 & Na & 1.2\\
Th & 0.0016 & K & 2.2\\
$^{40}$K & 0.00034 & Ti & 0.34\\
Si & 31.0 & P & 0.079\\
O & 47.8 & Mn & 0.023\\
Al & 9.6 & Mo & 0.002\\
Fe & 3.8 & H & 0.028\\
Ca & 1.3 & S & 0.3\\
Mg & 0.83 & Others & $<1.5$\\
\hline
\end{tabular}
\end{table}
The neutron yield $Y_{i}(E_{n})$ from ($\alpha , n$) reactions for a thick target $i$, have been taken from \cite{Mei}. The total neutron yield from rock is calculated by adding the individual neutron yield from the elements weighted by their mass ratio in the rock. The rock composition information given in Table~\ref{tab:table1} is used for this calculation\footnote{An intuitive program developed by the authors of Ref. \cite{Mei} available at \url{http://neutronyield.usd.edu/} has been used to obtain the neutron yield by providing the Jaduguda rock composition as input.}. The neutron yields from ($\alpha,n$) reactions in the rock have been obtained as $6.77\pm1.12$ yr$^{-1}$g$^{-1}$ from $^{238}$U and $5.33\pm0.90$ yr$^{-1}$g$^{-1}$ from $^{232}$Th.

\begin{figure}[h]
\centering
\includegraphics[width=0.6\textwidth]{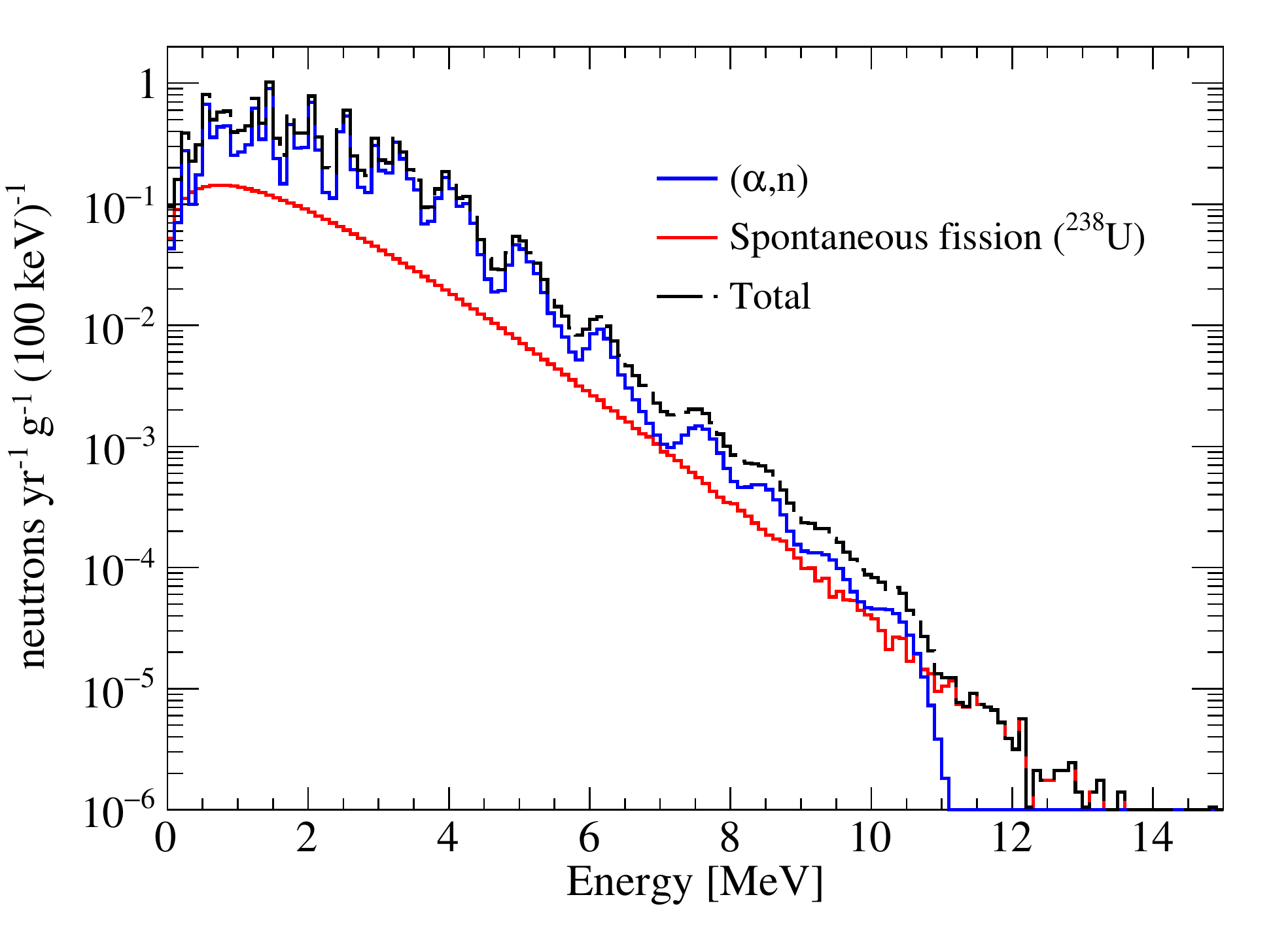} 
\caption{\label{fig:anflux}Expected specific neutron activity of Jaduguda rock due to ($\alpha,n$) reactions and spontaneous fission of $^{238}$U.} 

\end{figure}

The specific neutron activity due ($\alpha,n$) reaction in 1 g of rock normalized to the U and Th content is given in Figure ~\ref{fig:anflux}. It can be noticed that the neutrons from ($\alpha,n$) reactions have energies less than 12 MeV.

The natural abundance of $^{238}$U is 99.27\% with a spontaneous fission half-life of $8.2\times10^{15}$ years \cite{nubase,thiel}. The spontaneous fission half-life of $^{232}$Th is around 6 orders of magnitude higher and can be neglected \cite{nubase}. To simulate neutrons due to spontaneous fission, the Watt function has been used. The Watt function was initially used to explain fission due to thermal neutrons in $^{235}$U \cite{watt}. But it holds good for spontaneous fission of other heavy nuclei as well. The Watt function is given as
  \begin{equation}
  W(a,b,E^\prime)=a\sqrt{\frac{4a}{\pi b}}\exp\left(-\frac{b}{4a}-aE^\prime\right)\sinh(\sqrt{bE^\prime}),
  \end{equation}

where $a=1.54245$ MeV$^{-1}$ and $b=6.81057$ MeV$^{-1}$ are constants (for $^{238}$U) and $E^\prime$ is the secondary neutron energy \cite{verbeke}. The average neutron multiplicity ($\bar{\nu}$) for spontaneous fission of $^{238}$U is 2.01 \cite{verbeke}. Neutron yield due to spontaneous fission of $^{238}$U in the rock is estimated to be $3.43\pm0.55$ yr$^{-1}$g$^{-1}$.
\subsubsection{Transmission of radiogenic neutrons through rock}
\label{sub:alntrans}
The JUSL has a vertical rock overburden of 555 m. Radiogenic neutrons that are produced a few meters away from the rock-laboratory boundary, do not contribute to the neutron flux in the lab. A study is done to find the rock element thickness to be considered for the calculation. This will help in reducing statistical fluctuations and efficient calculation of neutron flux. The rock composition given in Table~\ref{tab:table1} is used for defining rock in GEANT4. The  radiogenic neutron energy distribution for simulation is generated using the specific neutron activity given in Figure~\ref{fig:anflux}. To calculate the neutron transmission probability, $10^5$ neutrons are randomly generated on the surface of a rock slab in an area 0.5 m $\times$ 0.5 m with momentum along $-Z$ direction. Thickness of the rock is varied and the length and breadth are fixed to 1 m. 
\begin{figure}[h]
\centering
\includegraphics[width=0.5\textwidth]{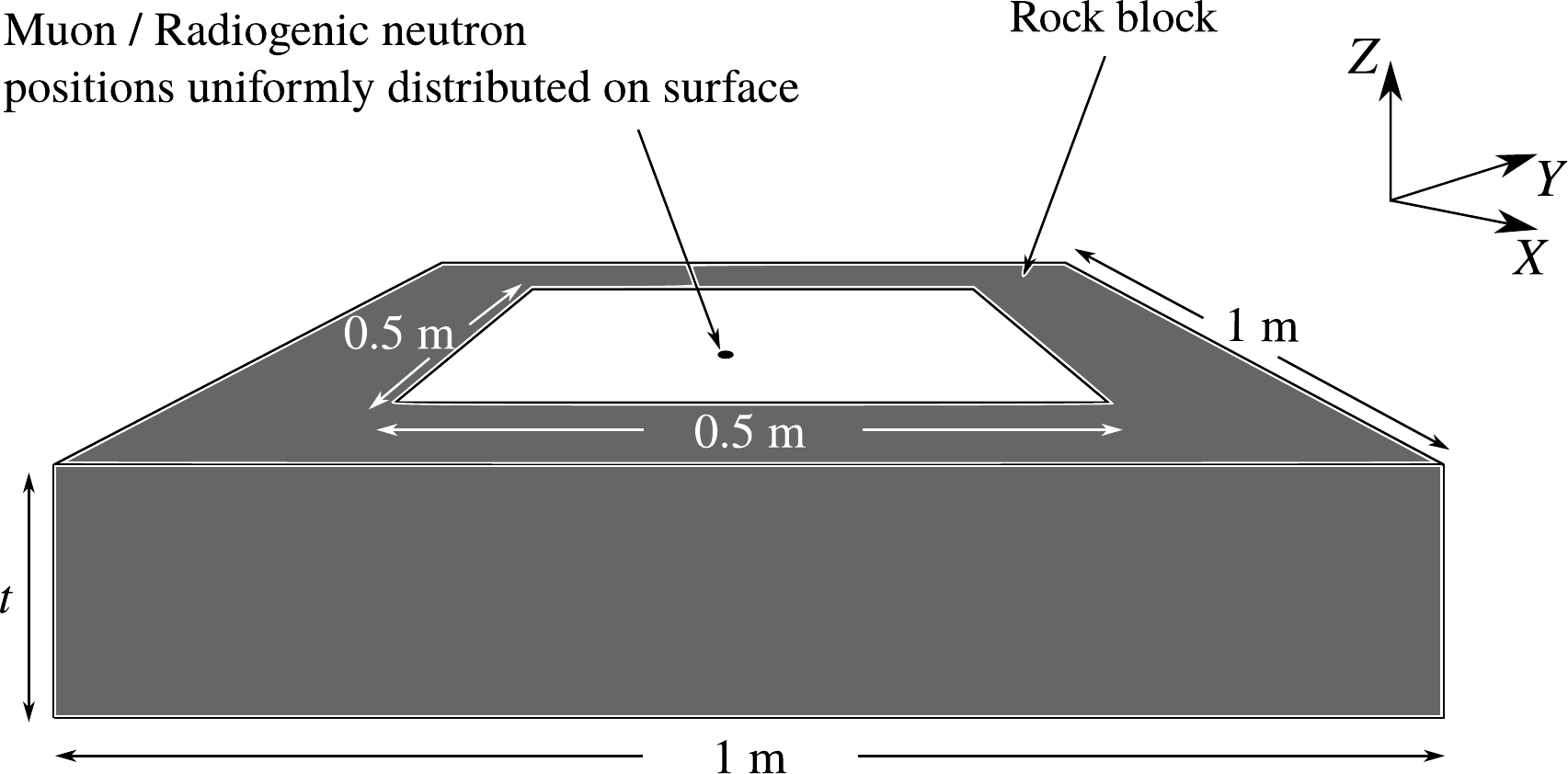} 
\caption{The rock slab model used in GEANT4 to calculate the transmission probability of neutrons.}
\label{fig:radNtrans}
\end{figure}

\begin{figure}[h]
\centering
\includegraphics[width=0.6\textwidth]{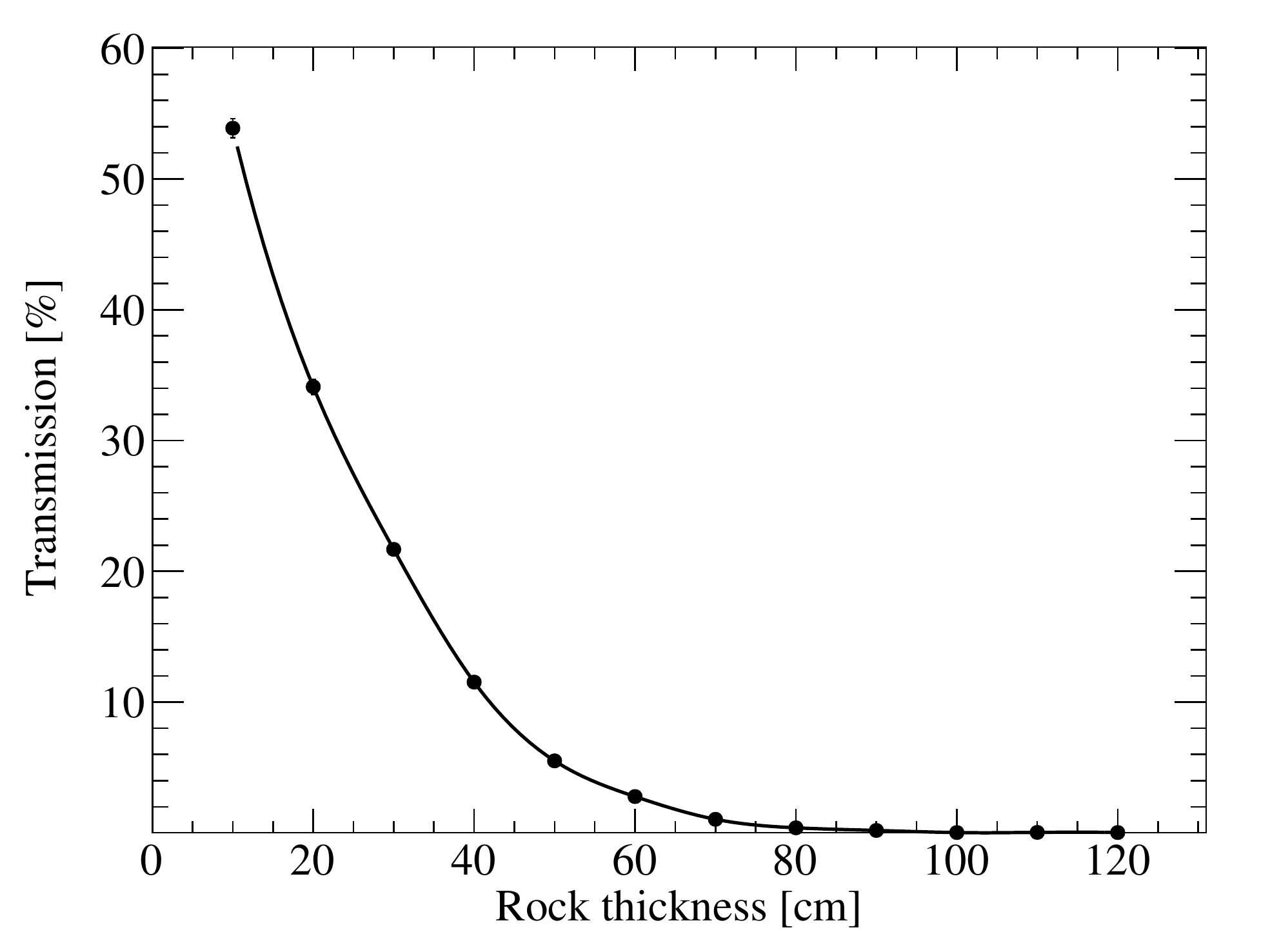}
\caption{Radiogenic neutron transmission probability as a function of rock thickness.}
\label{fig:TRradN}
\end{figure}

As neutrons propagate through the rock, they lose energy and get absorbed or scatter off. Neutrons coming out of the other side of the rock are recorded. The transmission probability, which is the ratio of the number of output neutrons to the number of input neutrons is calculated. It is shown as a function of rock thickness in Figure~\ref{fig:TRradN}.
Only about 0.5\% of total neutrons have been transmitted through rock of thickness 1.0 m. Therefore, for our simulation estimates, we consider 1 m of rock thickness surrounding the laboratory cavern as active material contributing to the radiogenic neutron flux.

\subsubsection{Flux of radiogenic neutrons in JUSL}
\label{sec:radflux}
\begin{figure}[h]
\centering
\subfigure[\label{alneve:a}]{\includegraphics[width=0.35\textwidth]{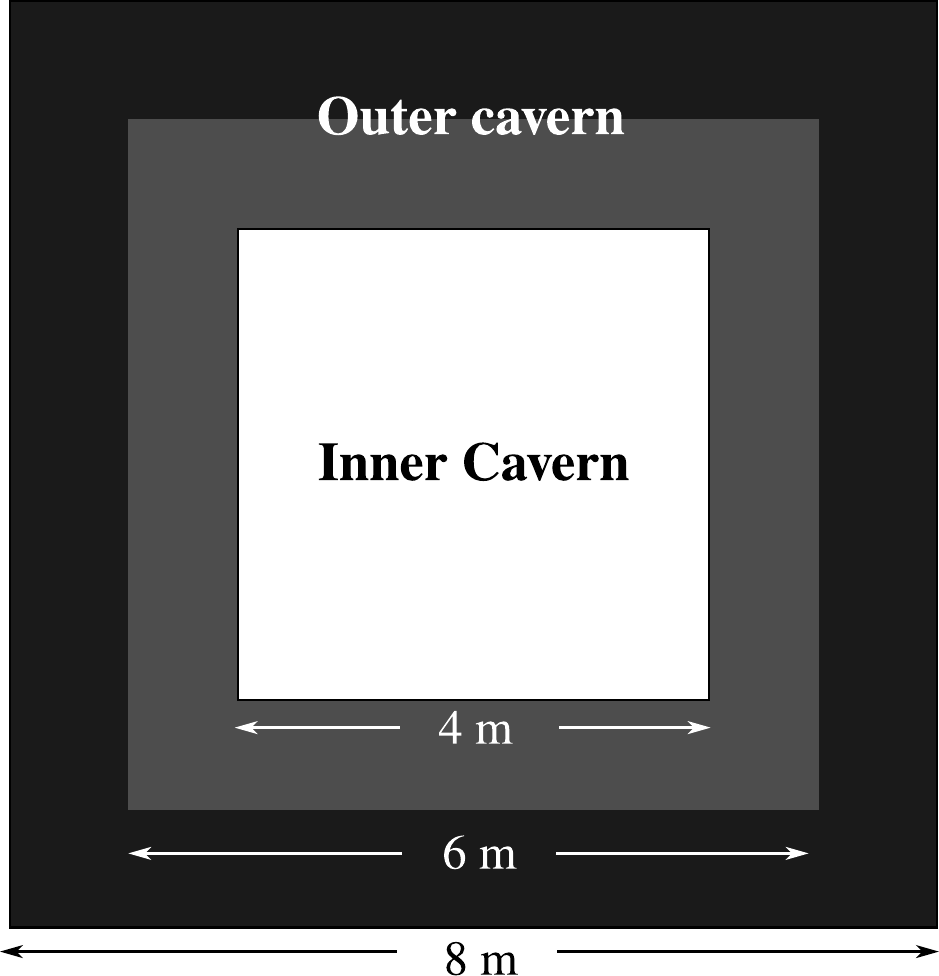}}
\subfigure[\label{alneve:b}]{\includegraphics[width=0.5\textwidth]{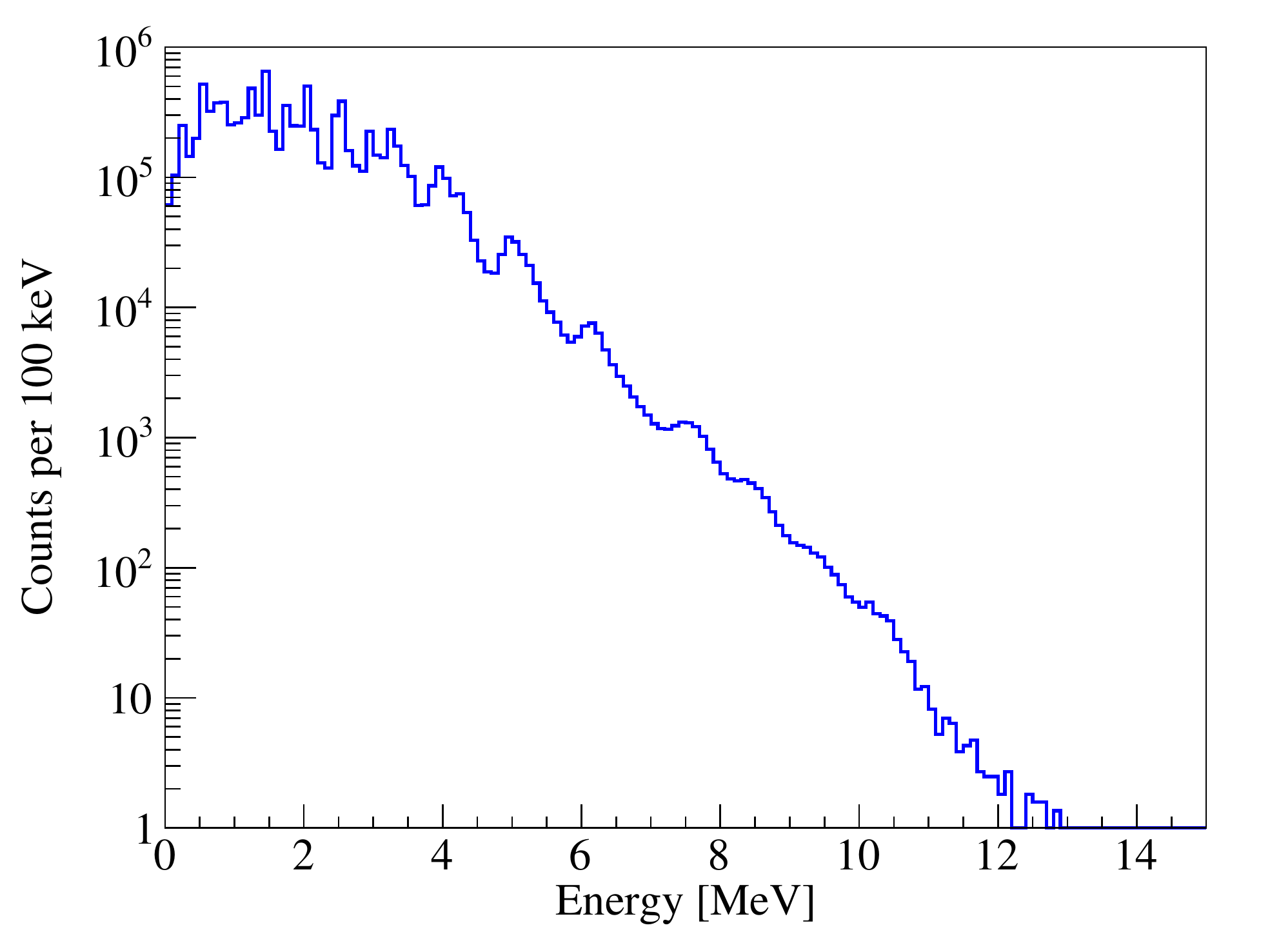}}
\caption{(a) The side view schematic of the cavern as implemented in GEANT4 to calculate the radiogenic neutron flux. (b) Energy distribution of radiogenic neutrons generated for $10^7$ events using the specific neutron activity shown in Figure \ref{fig:anflux}.}
\label{fig:alneve}
\end{figure}

We consider the cavern to be a hollow cube of outer side 8 m and inner side 4 m. In the cavern, the volume made up of rock is called the `Outer cavern' and the hollow volume is called the `Inner cavern'. To estimate the neutron flux in the laboratory, contributions from ($\alpha,n$) reactions and spontaneous fission are taken into account. In the rock, neutron events with the total energy distribution given in Figure~\ref{fig:anflux} and isotropic angular distribution, are generated in the 1 m thick grey region of the Inner cavern as shown in Figure~\ref{fig:alneve}(a). The energy distribution of 10$^7$ neutron events is shown in Figure~\ref{fig:alneve}(b). Some neutrons propagate through rock and some of them reach the experimental setup. These neutrons while propagating through rock can further produce neutrons.

\begin{figure}[h]
\centering
\includegraphics[width=0.6\textwidth]{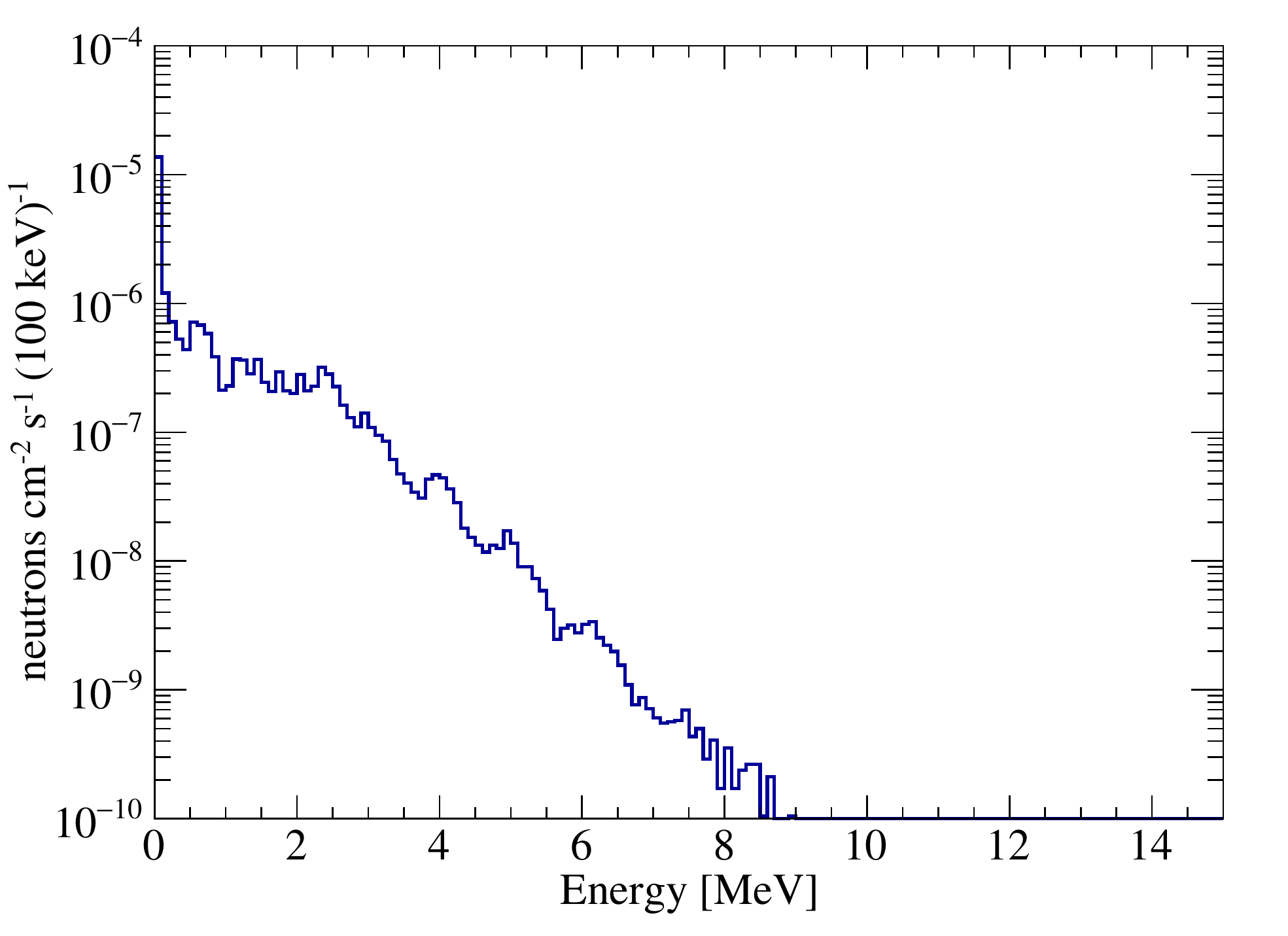}
\caption{Flux of radiogenic neutrons reaching laboratory shown as a function of energy.}
\label{fig:alnflux}
\end{figure}

Neutrons reaching the Inner cavern are recorded. The flux of radiogenic neutrons reaching the laboratory is obtained as 1.12($\pm0.11)\times 10^{-5}\thinspace$cm$^{-2}\thinspace$s$^{-1}$ above 100 keV (mean energy of 1.34 MeV) and 5.75($\pm0.58)\times 10^{-6}\thinspace$cm$^{-2}\thinspace$s$^{-1}$ above 1 MeV (mean energy of 2.18 MeV). 
The uncertainty shown in the parentheses include both statistical and systematic uncertainties. The same convention is followed throughout the paper unless otherwise specified.
The systematic uncertainty arises due to the variation of rock density and is around 10 \%. The flux of radiogenic neutrons reaching the laboratory as a function of energy is shown in Figure ~\ref{fig:alnflux}.

\subsection{Simulation of cosmogenic neutron flux}
\label{sec:cosmo}
Cosmic ray muons are created in the upper atmosphere from the interaction of primary cosmic rays with the atmospheric nuclei. They are the most abundant charged particles at sea level with an average energy of about 4 GeV and intensity $\sim$1 cm$^{-2}\thinspace$min$^{-1}$ \cite{pdg2018}. Since they are minimum ionizing particles, they can penetrate the rock overburden and reach the cavern. Cosmic ray muon events can be tagged using active veto shielding, but stopping them completely is not possible. Even though the flux of cosmic muons is small in underground laboratories, high energy muons can penetrate the rock and shielding materials and generate hadrons. Low energy muons get stopped in the rock.
\subsubsection{Cosmic muon event generation\label{sec:mugen}}
To generate realistic neutron flux, the muon flux at the experimental site needs to be determined. The minimum energy of muons required to reach the cavern and their maximum lateral displacement need to be calculated.

\subsubsection{Muon lateral displacement and maximum distance}
\begin{figure}[h]
  \begin{minipage}{0.5\linewidth}
    \includegraphics[width=\linewidth]{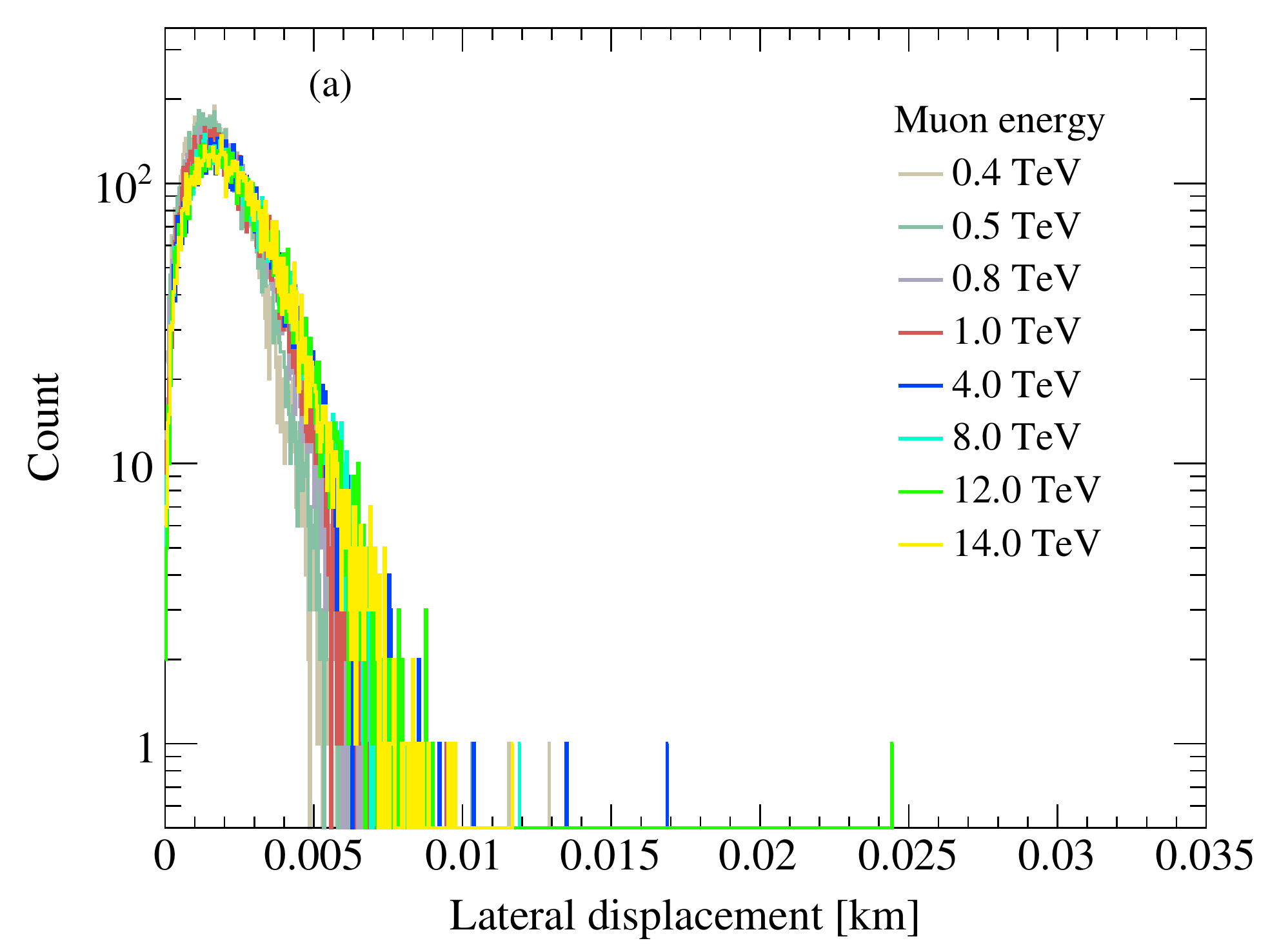}
  \end{minipage}
  \begin{minipage}{0.5\linewidth}
    \includegraphics[width=\linewidth]{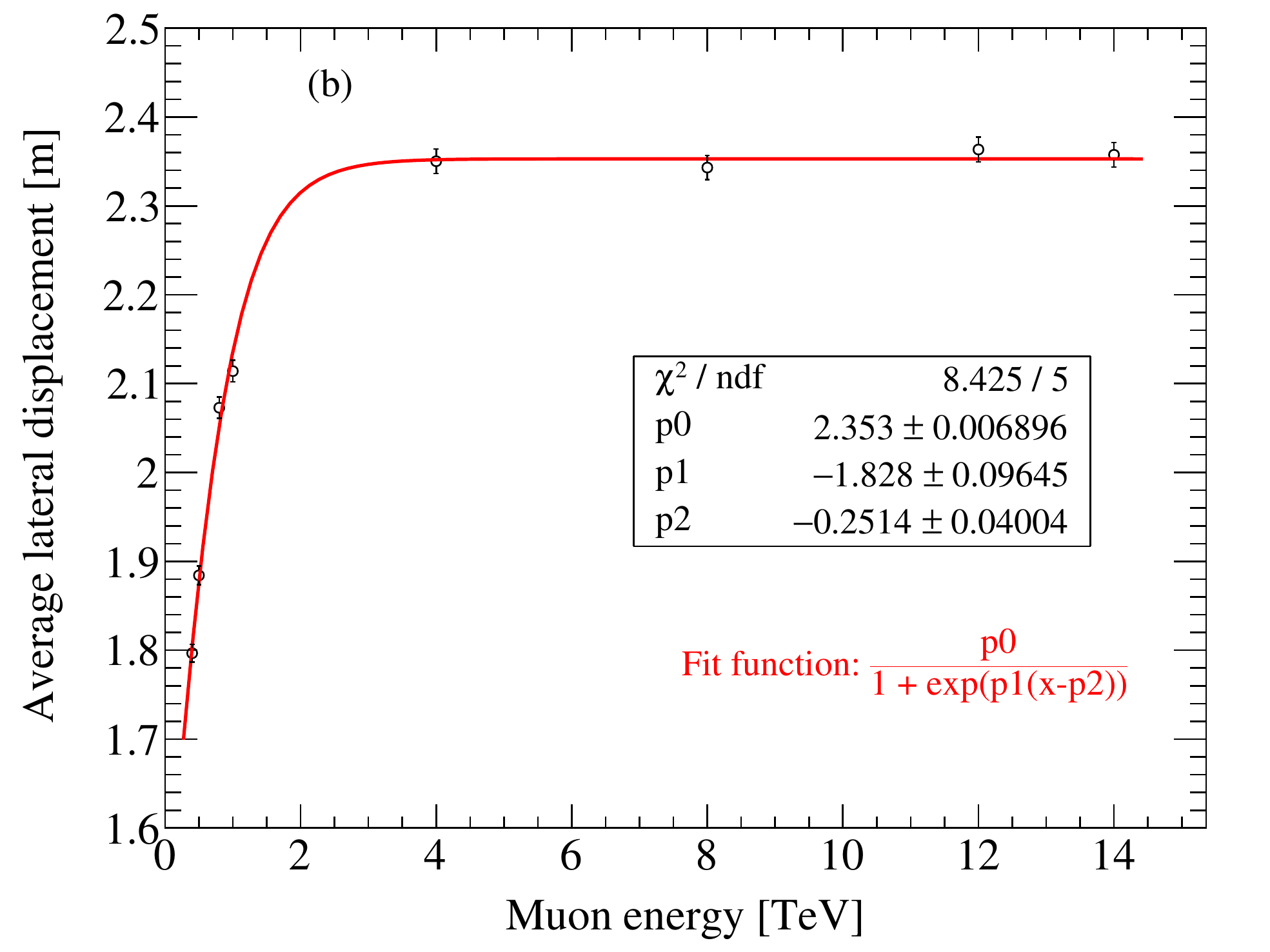}
  \end{minipage}\\
  \begin{minipage}{0.5\linewidth}
    \includegraphics[width=\linewidth]{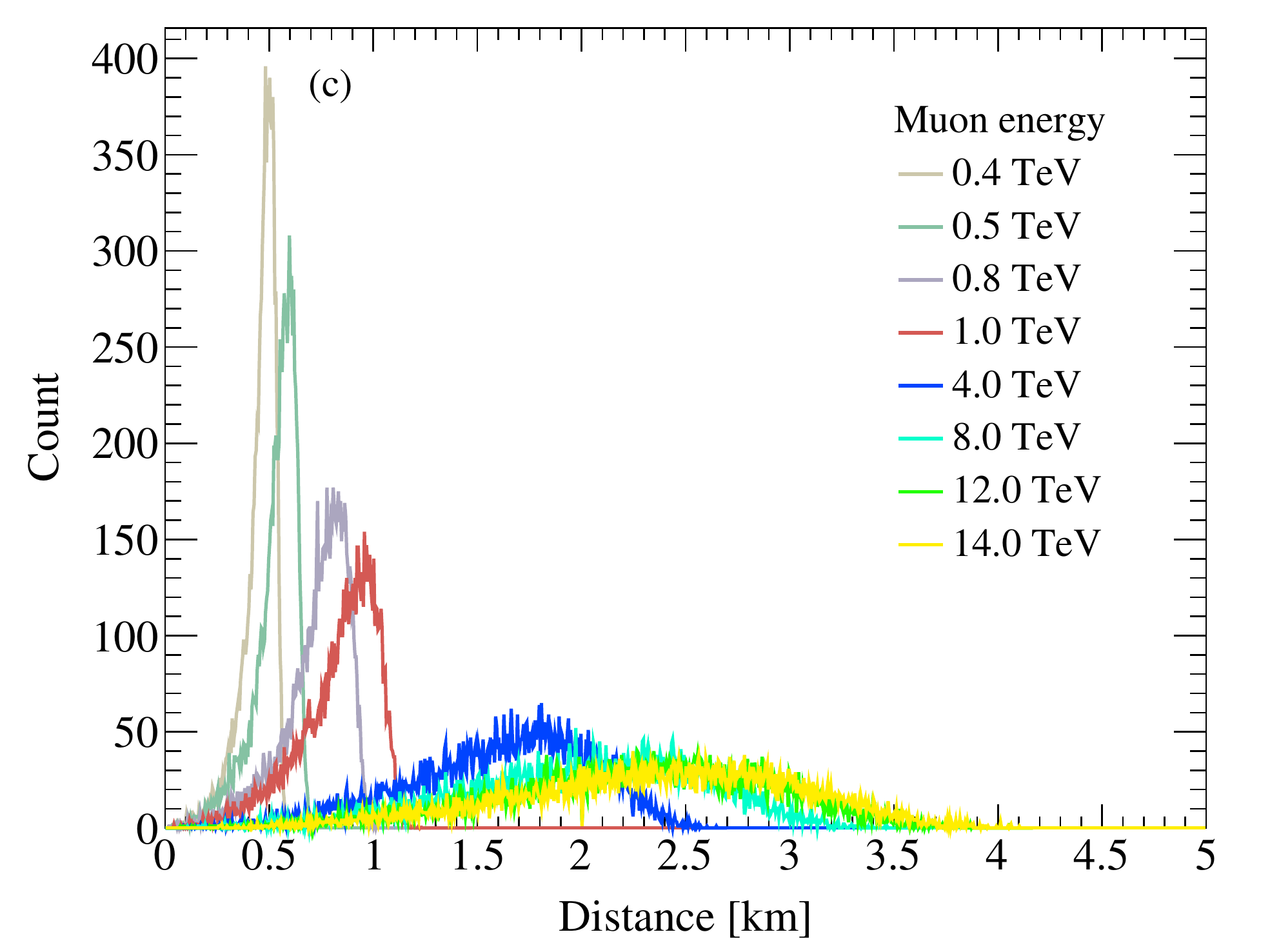}
  \end{minipage}
  \begin{minipage}{0.5\linewidth}
    \includegraphics[width=\linewidth]{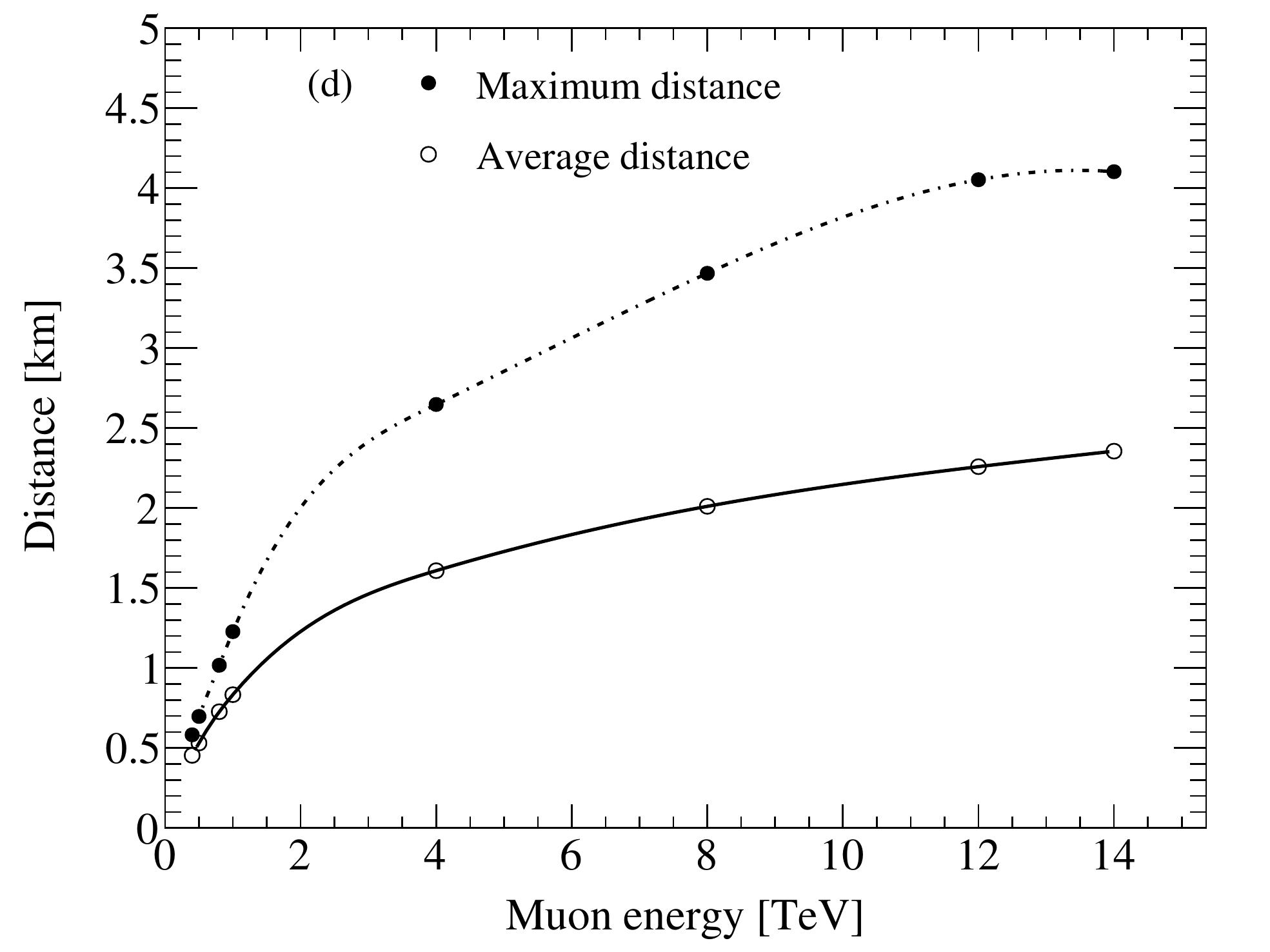}
  \end{minipage}
  \caption{\label{fig:latdisp}(a) Lateral displacement distribution of muons from their initial direction of propagation in the rock. (b) Average lateral displacement as a function of muon energy. (c) Distribution of distance traversed in the rock by muons of different incident energies. (d) Maximum and average distance traversed by muons as a function of energy.}
\end{figure}

To reach the cavern from the surface, cosmic muons have to be quite energetic. Moreover, muons interact with the earth/rock and undergo scattering. Their initial direction of propagation is altered and their expected position at a depth is displaced. By calculating the maximum distance traversed by muons of given energy, we can estimate the minimum energy required by muons to reach the cavern. Muons with different fixed energies were made to pass through a cube of rock of side 6 km along the $-Z$ direction. The maximum distance traversed and lateral displacement were calculated. The results are shown in Figure \ref{fig:latdisp}(a-d). From Figures \ref{fig:latdisp}(a) and \ref{fig:latdisp}(b), it can be seen that the average lateral displacement saturates to $\sim2.3$ m. But the maximum lateral displacement can be as high as 30 m. From Figures \ref{fig:latdisp}(c) and \ref{fig:latdisp}(d) we see that the minimum energy of muon that can reach the cavern (depth 555 m) is around 300 GeV. Therefore, we do not consider cosmic muons of energy less than 300 GeV in our simulation. The lateral displacement helps in making the simulation faster and is discussed in the next section.
\subsubsection{Calculation of muon flux at the cavern}
\begin{figure}[h]
  \centering
  \includegraphics[width=0.65\linewidth]{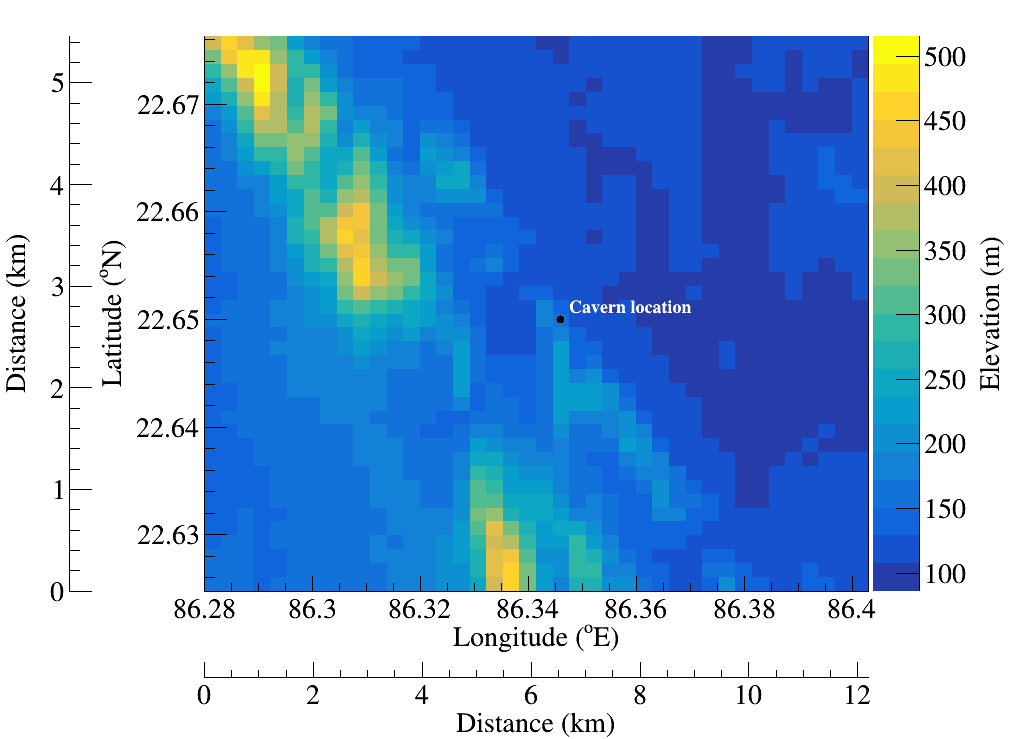}
\caption{Elevation map of the area around JUSL (12.2 km $\times$ 5.45 km).}
\label{fig:jmap}
\end{figure}

The latitude, longitude and elevation information of Jaduguda area has been obtained using Google Earth Pro \cite{jgmap} and the topological profile is shown in Figure ~\ref{fig:jmap}. The density of rock is taken to be 2.89~g$\thinspace$cm$^{-3}$ \cite{JR}.
\begin{figure}[h]
  \centering
  \includegraphics[width=0.65\linewidth]{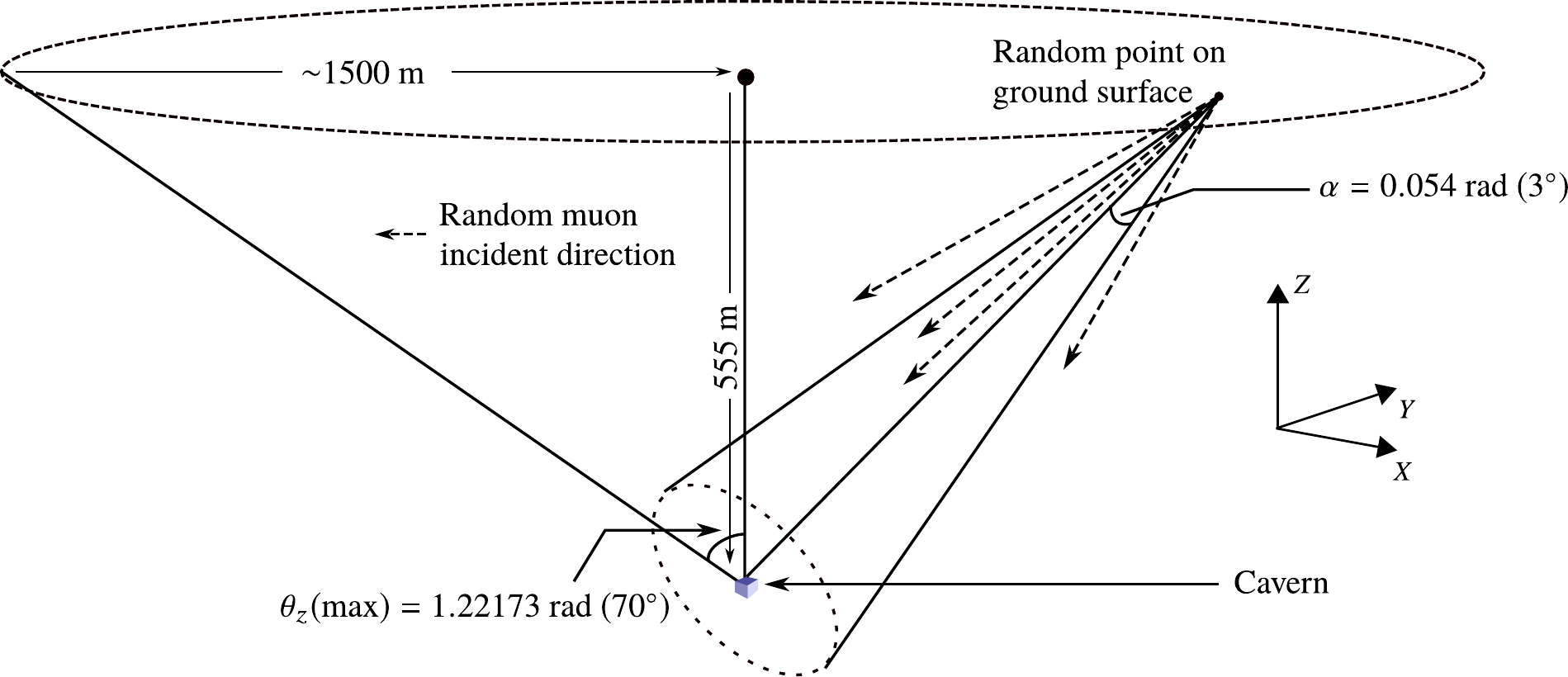}
\caption{Schematic describing the methodology of calculating muon flux at the cavern.}
\label{fig:jmap_strat}
\end{figure}

The energy and angular distribution of muons on the ground surface are generated using Gaisser's parameterization \cite{gaisser}
\begin{equation}
\label{eq:f0}
\frac{\mathrm{d}^2N_{\mu}}{\mathrm{d}E_{\mu}\mathrm{d}\Omega} \approx \frac{0.14
E_{\mu}^{-2,7}}{\hbox{cm}^2\thinspace\hbox{s}\thinspace\hbox{sr}\thinspace\hbox{GeV}}
 \times \left[\dfrac{1}{1 + \dfrac{1.1 E_{\mu} \cos\theta_z}{\epsilon_\pi}} +
\dfrac{\eta}{1 + \dfrac{1.1 E_{\mu}\cos\theta_z}{\epsilon_K}}\right],
\end{equation}
where $\theta_z$ is the zenith angle, $E_{\mu}$ is the energy of the muon, $N_\mu$ is the number of muons, $\Omega$ is the solid angle, $\epsilon_\pi$ = 115 GeV, $\epsilon_K$ = 850 GeV and $\eta$ = 0.054. 

This parametrization is valid when the decay of muon is negligible and curvature of the earth can be neglected (i.e. for zenith angle $\theta_z < 70^\circ$). The azimuthal angle $\phi$ distribution is generated uniformly. Muons are propagated to the experimental hall/cavern at 555 m below the surface. As muon propagates through the earth/rock, it scatters and loses energy.

\begin{figure}[h]
\includegraphics[width=0.50\linewidth]{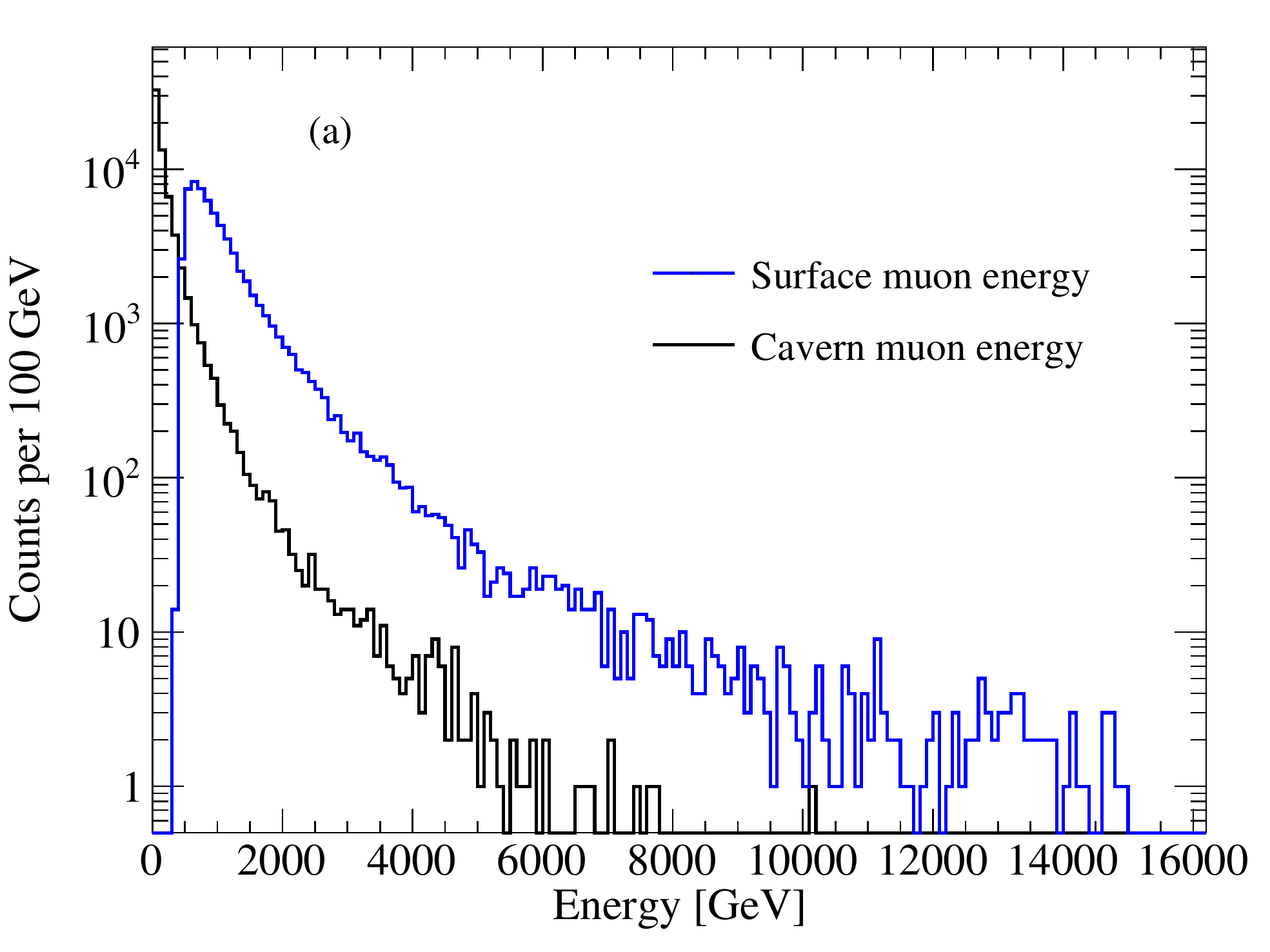}
\includegraphics[width=0.50\textwidth]{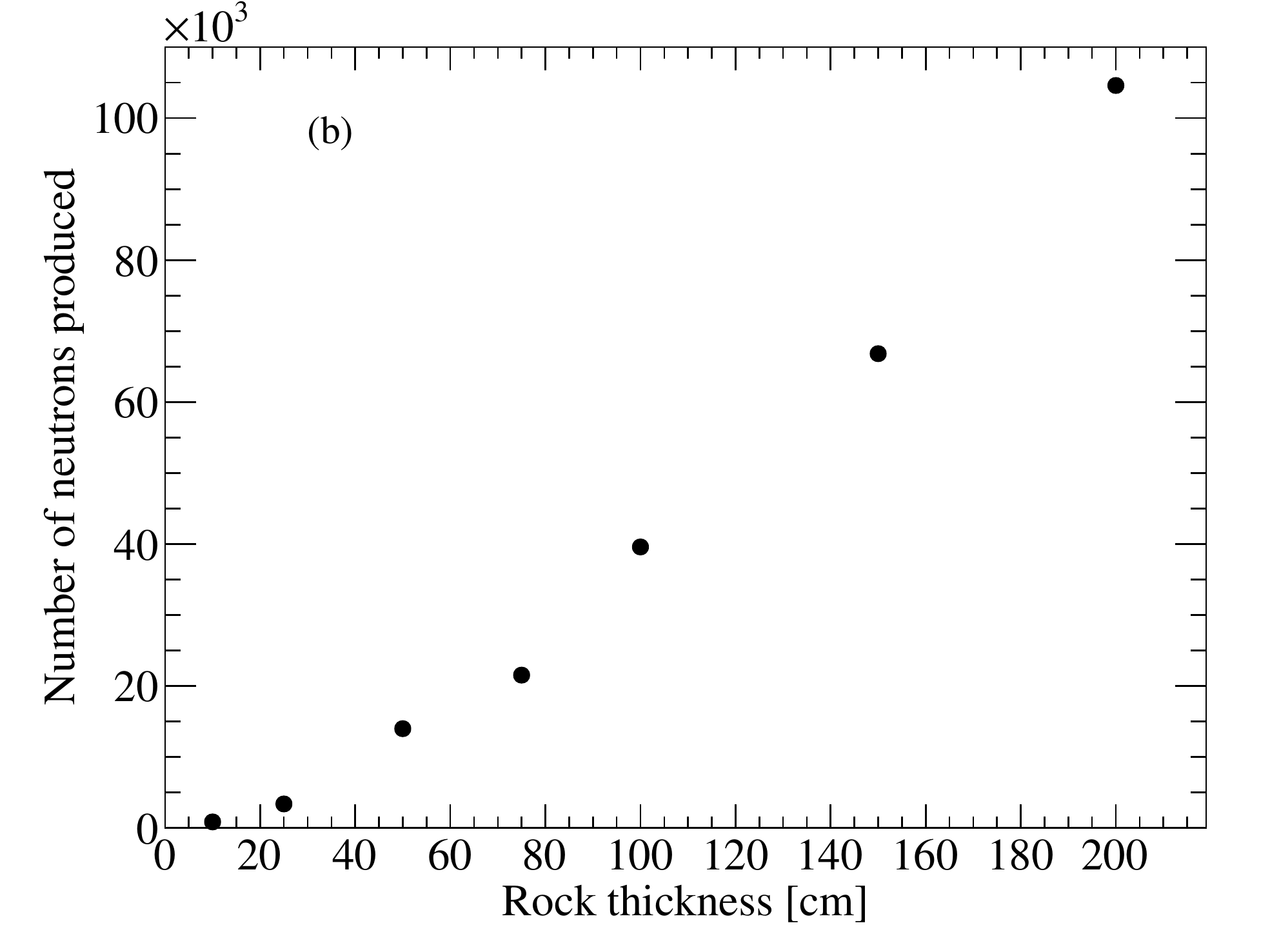}
\includegraphics[width=0.50\textwidth]{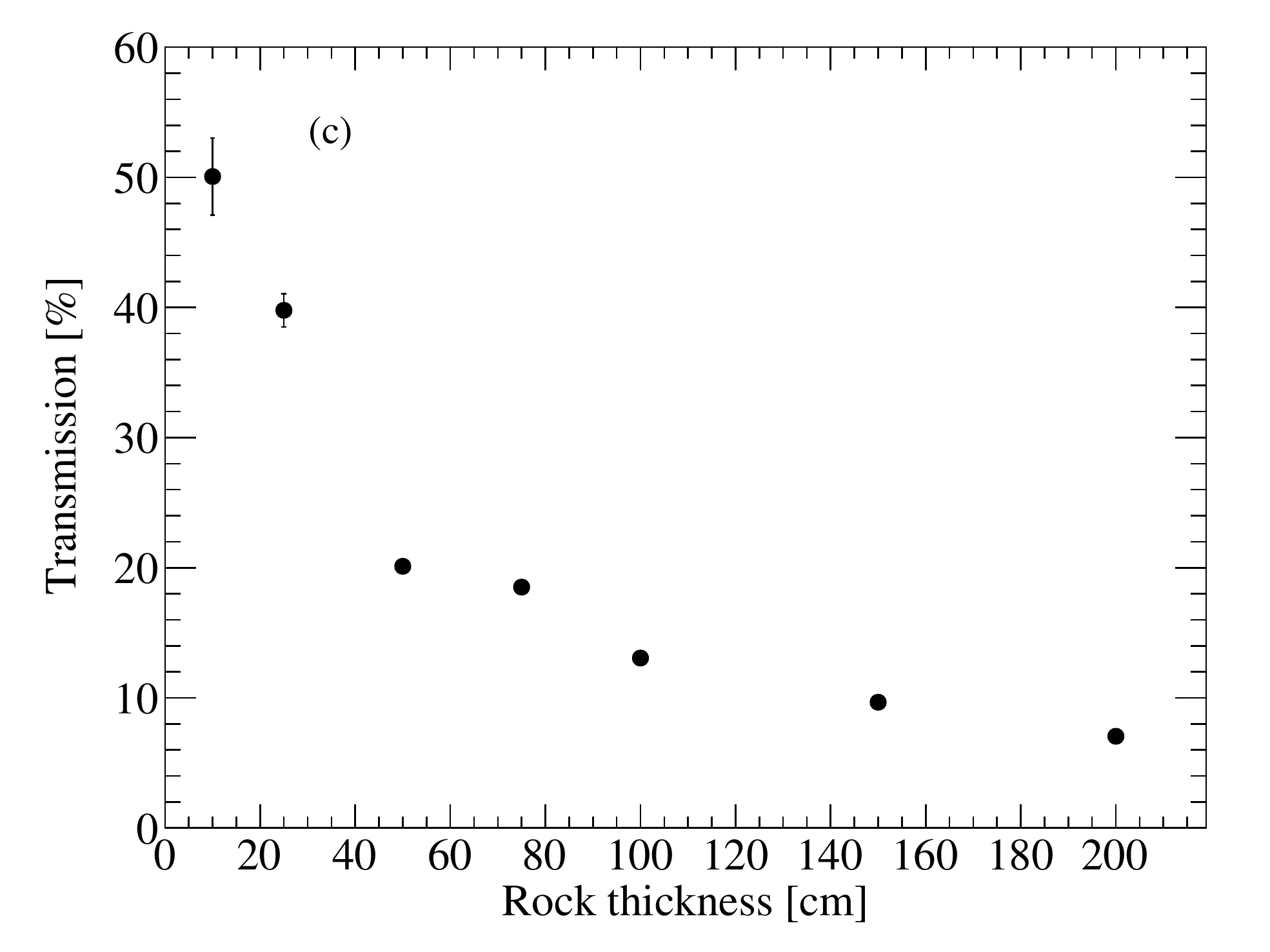}
\includegraphics[width=0.50\textwidth]{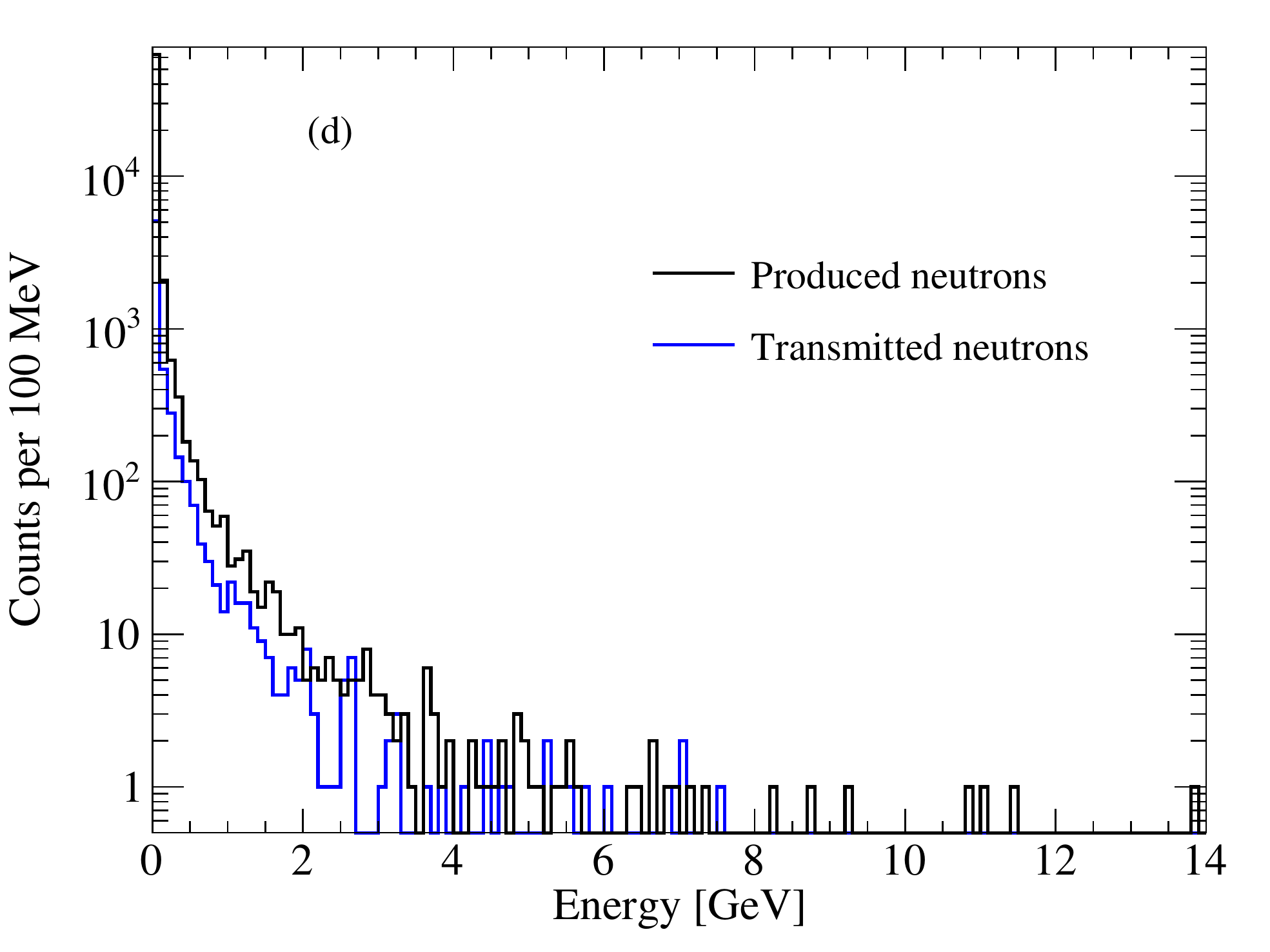}
\caption{(a) Energy distribution of muons at the surface and after reaching the cavern. (b) Neutron produced from muon interaction as a function of rock thickness. (c) Neutron transmission probability as a function of rock thickness. (d) The spectra of neutron energy at production ($N_\mathrm{prod}$) and after transmission through rock of 150 cm thickness ($N_\mathrm{out}$) shown upto 14 GeV.}
\label{fig:munprod}
\end{figure}

Muon events are generated in the range 330 GeV to 15 TeV. Both $\mu^{+}$ and $\mu^{-}$ events are generated. Since muons impinging on the surface at $\theta_z < 70^\circ$ are considered, they are generated over a circular area of radius $\sim$1.5 km with the center at cavern location as shown in Figure \ref{fig:jmap_strat}. Altitude information ($z$) at each point $(x,y)$ is obtained from the topological profile (Figure \ref{fig:jmap}). The altitude dependence of muon flux is ignored. Since a huge number of events\footnote{The number of muons that are expected to fall over a circular area of radius 1.5 km in a day is $\sim20\times10^9$.} are required to be generated on the surface to  have a reasonable number of muons reaching the cavern, the simulation becomes time consuming and computationally intensive. In order to avoid laborious simulation, we can use the lateral displacement mentioned in the previous section to save time. From Figure \ref{fig:latdisp}(a), as the maximum lateral displacement is about 30 m, muons with incident direction within a cone having an axis as the line connecting the point of incidence to the center of the cavern and radius 30 m (cone opening angle of $\alpha\sim3.1^{\circ}$) are simulated (See Figure \ref{fig:jmap_strat}). The other muons are not simulated and are counted as incident but not reaching the cavern. The energy distribution of muons before and after reaching the cavern surface by passing through the earth/rock is shown in Figure \ref{fig:munprod}(a). It is observed that the flux of muons on the top surface of the outer cavern is $4.49(\pm0.25)\times10^{-7}\thinspace$cm$^{-2}\thinspace$s$^{-1}$. The information of muons such as position, momentum etc. are recorded.

Since the energy of neutrons produced from muon interaction gets attenuated in the rock, similar to the case of radiogenic neutrons, only a rock element with finite size will contribute to the flux of cosmogenic neutron in the laboratory. A study similar to that described in the section \ref{sub:alntrans} is done for finding the rock thickness to be considered for simulating the muon induced neutron flux in the experimental hall.

The rock geometry given in Figure~\ref{fig:radNtrans} is used for the simulation. The energy distribution of muons at the surface of the outer cavern shown in Figure~\ref{fig:munprod} (blue histogram) is used. The average energy of muons is around 200 GeV. Muons are propagated from random positions on a plane of dimension (0.5 m $\times$ 0.5 m) (Figure ~\ref{fig:radNtrans}(a)) through the rock in the $-Z$ direction. The muon interactions with rock generate neutrons. The neutrons coming out on the other side of the rock are recorded. The simulation is repeated for different rock thicknesses ($t$ = 10 cm, 25 cm, 50 cm, 75 cm, 100 cm, 150 cm).

Number of neutrons produced ($N_\mathrm{prod}$) as a function of rock thickness is given in Figure~\ref{fig:munprod}(b). As the rock thickness increases, the neutron production also increases due to the increase in the probability of interaction.
\subsubsection{Transmission of cosmogenic neutrons through rock}
Neutron energy spectra at production ($N_\mathrm{prod}$) and after transmission through rock ($N_\mathrm{out}$) of thickness 150 cm are shown in Figure~\ref{fig:munprod}(d). The production rate of neutrons for 150 cm rock thickness is around 0.1 neutron/muon.

The neutron transmission probability which is the ratio of number of neutrons coming out on the other side of the rock to the number of neutrons produced in the rock ($R = N_\mathrm{out}/N_\mathrm{prod}$), as a function of rock thickness is shown in Figure~\ref{fig:munprod}(c). As the rock thickness increases, the transmission probability decreases. About 7\% of total neutrons are transmitted through rock of thickness 200 cm.
\subsubsection{Flux of muon induced neutrons in JUSL}
\begin{figure}[h]
\begin{minipage}{0.5\linewidth}
\centering
\includegraphics[width=0.8\linewidth]{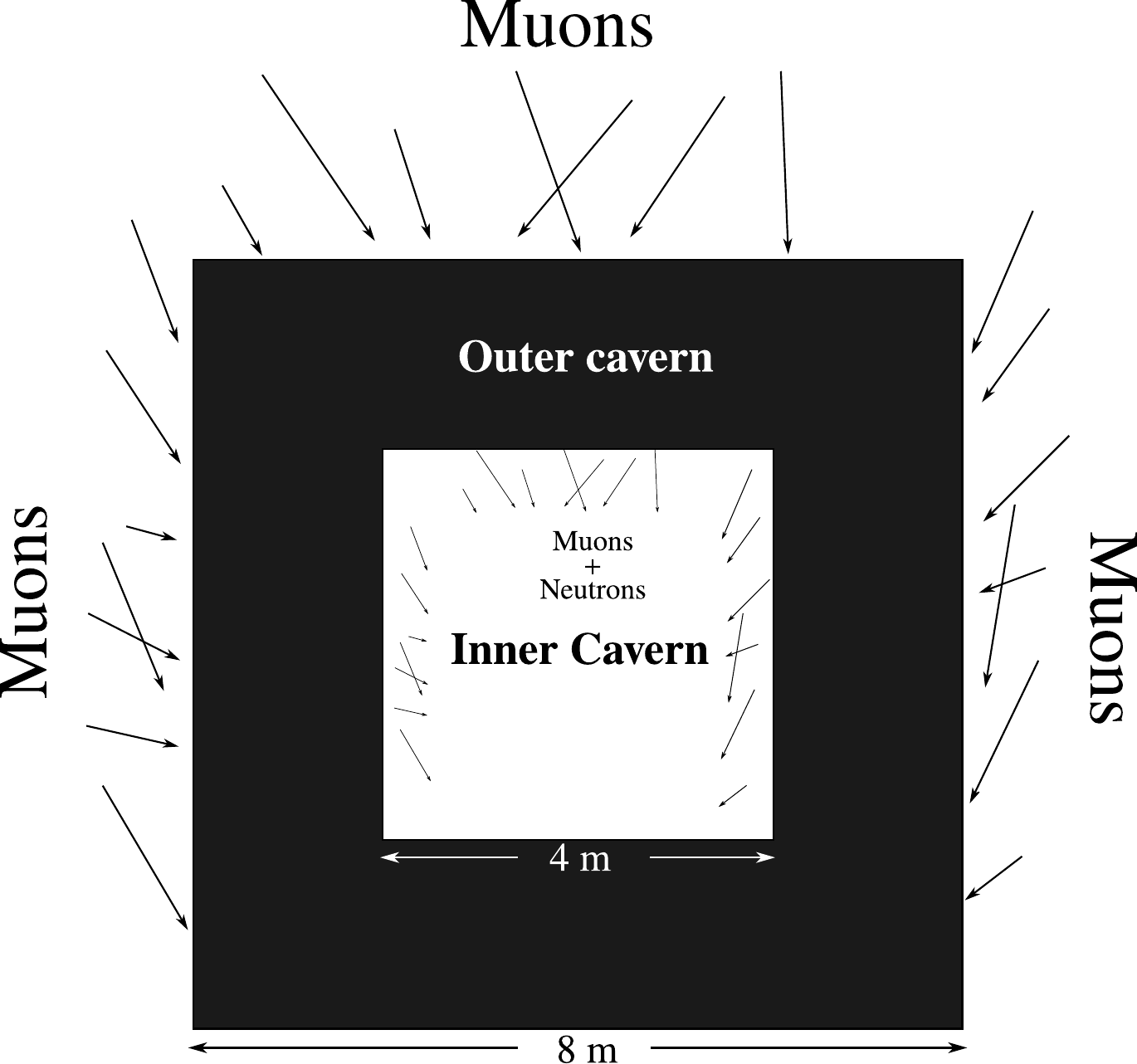}\\
\small{(a)}
\end{minipage}
\begin{minipage}{0.5\linewidth}
\centering
\includegraphics[width=\linewidth]{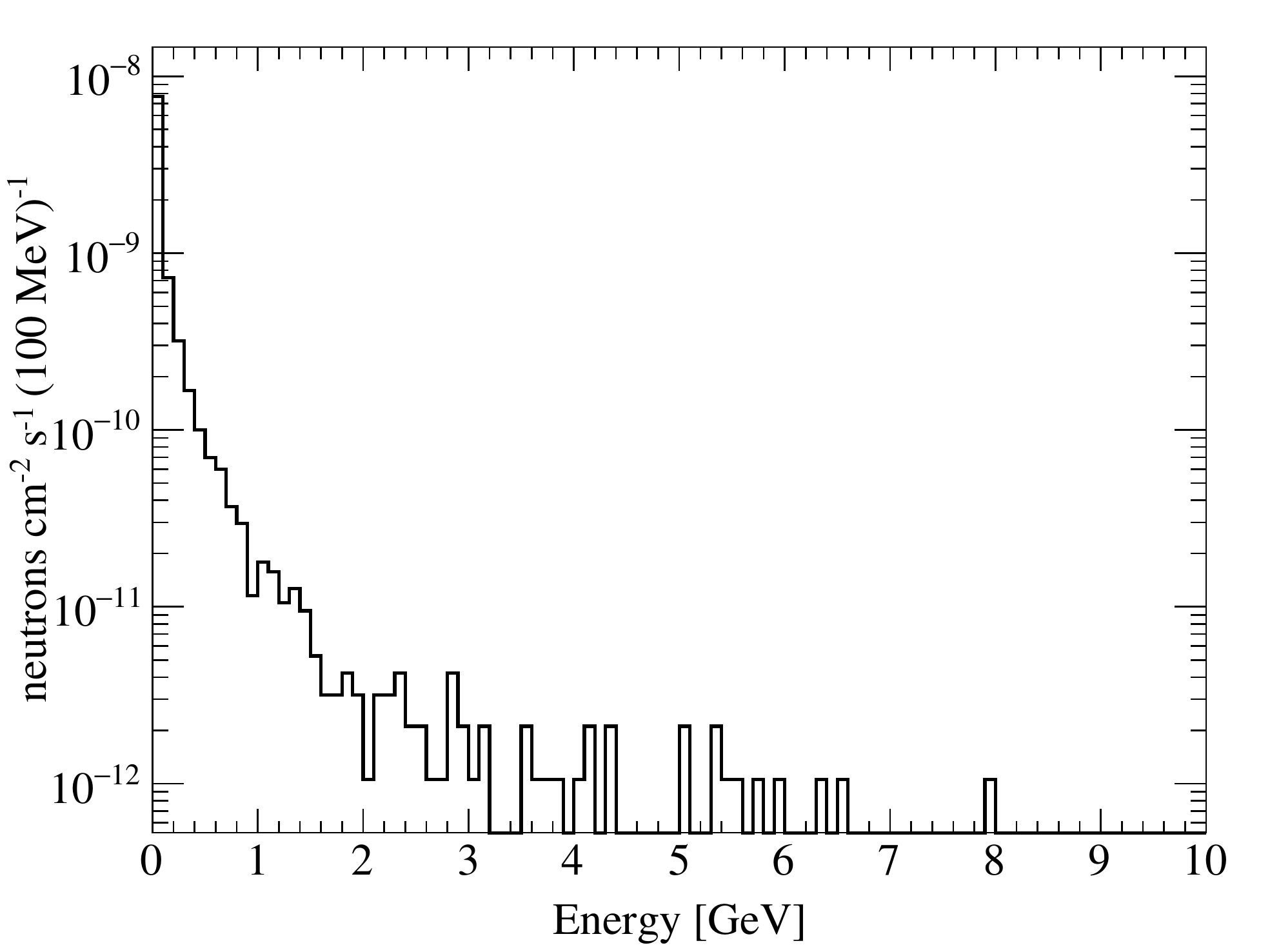}\\
\small{(b)}
\end{minipage}
\caption{(a) Geometry used for simulation. (b) Flux of muon induced neutrons in the Inner cavern.}
\label{fig:mungeo}
\end{figure}

To estimate the cosmogenic neutron background at the JUSL, the muon flux obtained at the surface of the outer cavern is used. The muon events are generated on the five surfaces of the 2 m thick rock around the cavern as shown in the Figure~\ref{fig:mungeo} (front and back surfaces are not shown in the figure). Muons are allowed to propagate through the rock and reach the cavern. There are no muons propagating from the bottom side. While going through the rock, they generate neutrons and other shower particles like hadrons, gamma and electrons which then enter the laboratory. Some of the neutrons get absorbed in the rock itself.

The flux of neutrons reaching the cavern is shown in Figure ~\ref{fig:mungeo}(b). It can be seen that the neutrons produced in rock have energies up to 10s of GeVs. The muon induced neutron flux in the cavern is found to be $0.93(\pm0.08)\times 10^{-8}\thinspace$cm$^{-2}\thinspace$ s$^{-1}$ with no energy threshold and $7.25(\pm0.65)\times 10^{-9}\thinspace$cm$^{-2}\thinspace$ s$^{-1}$ above 1 MeV. The systematic uncertainty is due to the variation of the rock density and the fact that around 7\% of neutrons can still come from distances greater than 2 m from the cavern boundary.

\subsection{Total neutron flux in JUSL}
\label{sec:totN}
\begin{figure}[h]
  \centering
  \includegraphics[width=0.6\linewidth]{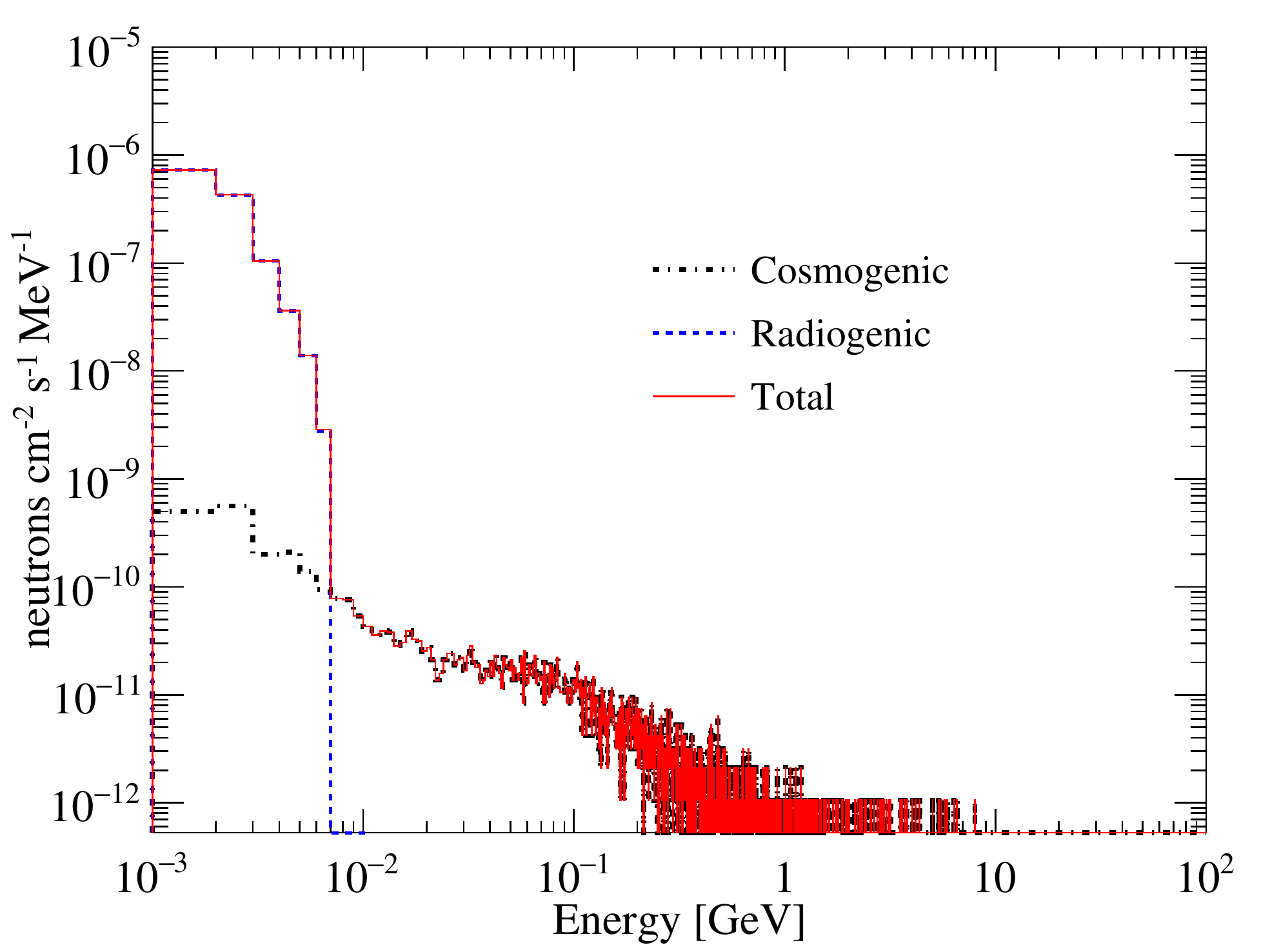}
\caption{Total neutron flux due to radiogenic and cosmogenic sources expected at the cavern shown as a function of energy.}
\label{fig:totnf}
\end{figure}
The total neutron flux from radiogenic origin and muon interactions reaching the laboratory in the energy range 1 MeV - 100 GeV is shown in the Figure ~\ref{fig:totnf}. Above 1 MeV energy threshold, the flux of radiogenic neutrons is 5.75$(\pm0.58)\times$ 10$^{-6}\thinspace$cm$^{-2}\thinspace$s$^{-1}$, whereas the flux of neutrons produced by muon interaction in rock is 7.25($\pm0.65)\times$ 10$^{-9}\thinspace$cm$\thinspace^{-2}\thinspace$s$^{-1}$. For energies less than $\sim$10 MeV, neutrons flux from radiogenic neutrons is around 3 orders of magnitude greater than the muon induced neutron flux. For energies above 10 MeV only muon induced neutrons contribute to the spectrum. Therefore the total neutron flux reaching the cavern/laboratory above 1 MeV energy threshold is found to be 5.76($\pm0.58)\times$ 10$^{-6}\thinspace$cm$\thinspace^{-2}\thinspace$s$^{-1}$. Our values are comparable with neutron flux estimations done for dark matter experiments at Boulby and WIPP salt mines \cite{boulby,WIPP}.

\section{Shielding combinations to reduce neutron flux\label{sec:shield}}
A typical experimental setup usually consists of several layers of active and passive shielding. Active shielding can veto muons and associated neutrons. Passive shielding systems consist of Lead(Pb) or Iron(Fe) for shielding gammas, hydrocarbons for moderating neutrons and copper for attenuation of gammas. 

For simplicity, a rectangular geometry of shielding ($x~=~$1 m, $y = $ 1 m for all the layers) is considered as shown in Figure ~\ref{fig:shield1}(a)(side view). There is an air gap of 1 cm between each layer. Thickness of the materials is varied. Different combinations of Pb and Polypropylene are studied to find the optimal shielding composition for neutron reduction.
\subsection{Reduction of radiogenic neutron flux}
Radiogenic neutrons reaching the cavern as obtained in Section~\ref{sec:radflux}, Figure \ref{fig:alnflux}  are allowed to pass through the shielding.
All the radiogenic neutrons are stopped by 40 cm thick polypropylene shielding and number of neutrons reaching Pb surface is found to be zero. Radiogenic neutrons generated from rock have energy upto a few MeVs and can be shielded using an outer polypropylene layer only.

\subsection{Reduction of muon induced neutron flux}

\begin{figure}[h]
\subfigure[\label{shield1:a}]{\includegraphics[width=0.45\textwidth]{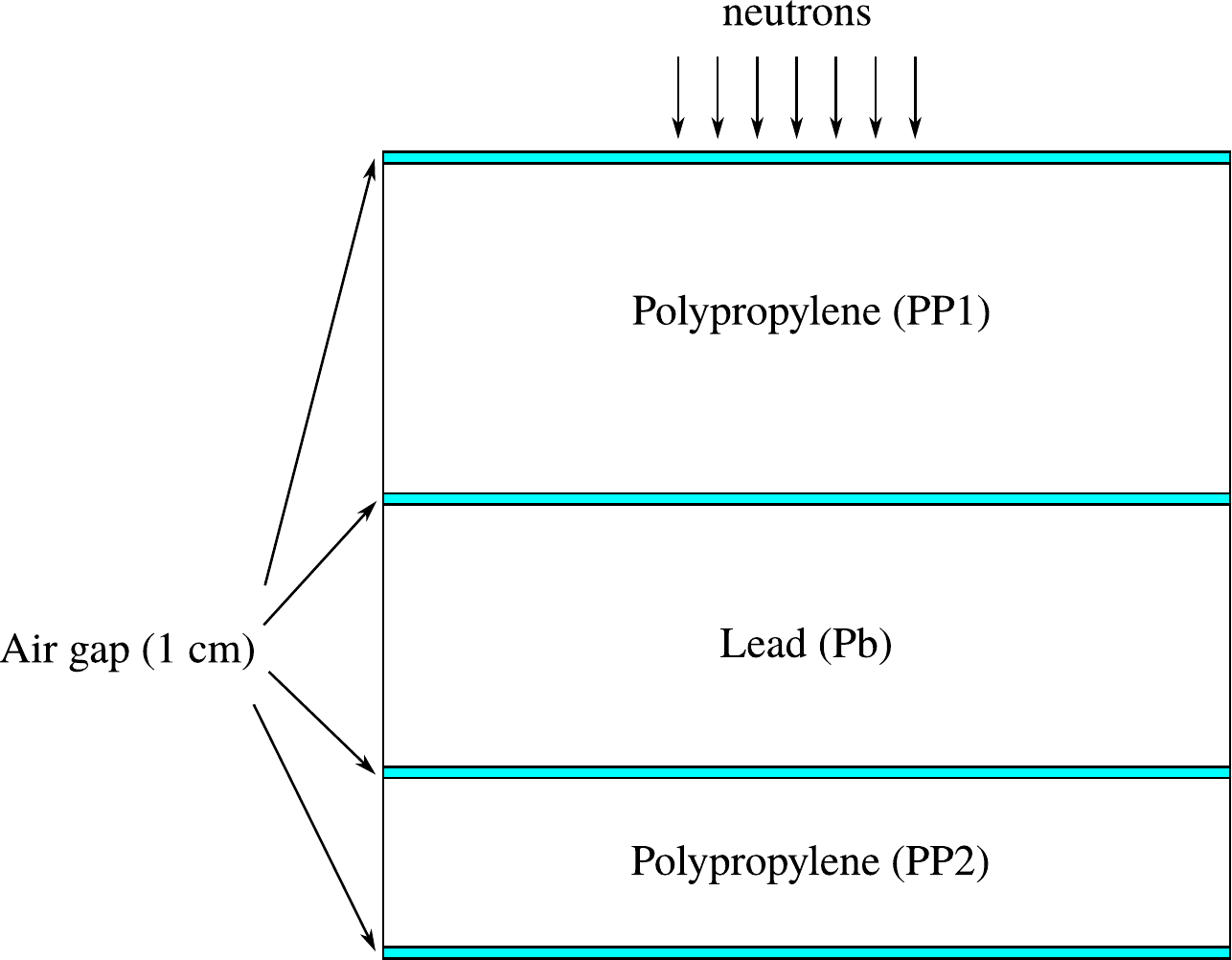}}
\subfigure[\label{shield1:b}]{\includegraphics[width=0.5\textwidth]{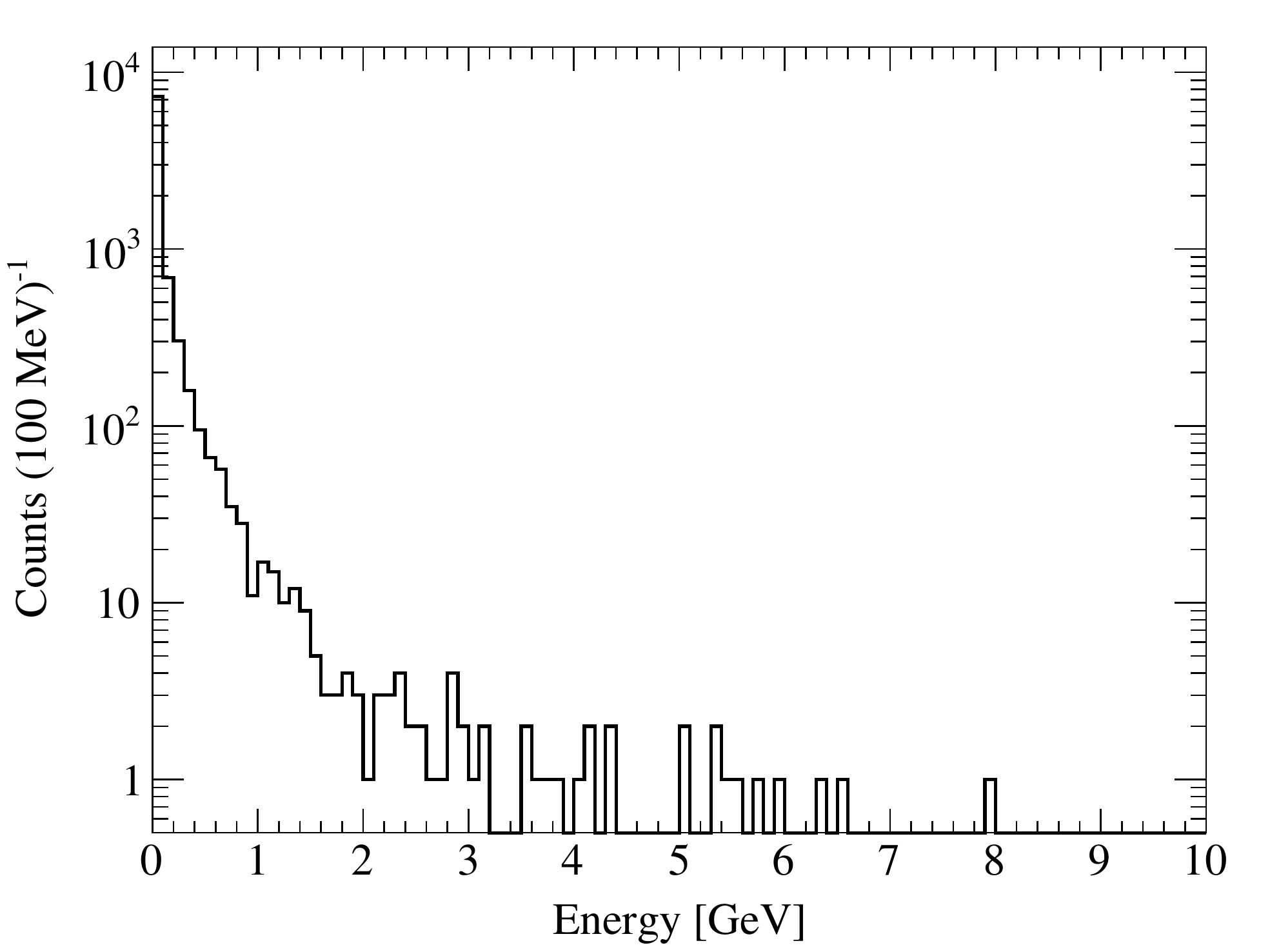}}
\caption{(a) Rectangular shielding layers used for simulation. The thicknesses of Pb and PP2 are varied. (b) Energy distribution of generated cosmogenic neutrons.}
\label{fig:shield1}
\end{figure}

The flux of neutrons produced from muon interactions in rock (Figure ~\ref{fig:shield1}(b)) is used to generate 10$^4$ neutrons on a plane of area (0.5 m $\times$ 0.5 m) which is at 1 cm above the polypropylene layer (PP1) of thickness 40 cm. Neutrons crossing the boundary of each layer are recorded. 

\begin{table}[h]
  \centering
    \caption{\label{tab:shielddes}Different shielding configurations and their effectiveness. Uncertainties shown are statistical only.}
  \begin{tabular}{c|p{1 cm}|p{1 cm}|p{1 cm}|c}
    \hline
    & \multicolumn{3}{c|}{Thicknesses of different}
    & \\
    Configuration  & \multicolumn{3}{c|}{shielding layers (cm)}
    & Transmission (\%)\\
    \cline{2-4}
    & \centering PP1 &  \centering Pb & \centering  PP2 
    & \\
    \hline
    CFG-1 & \centering 40	& \centering -- & \centering -- 
    &	$52.31	\pm	0.72$	\\
    CFG-2 & \centering 40 & \centering 30 & \centering -- 
    & $136.3	\pm	1.17$	\\
    CFG-3 & \centering 40 & \centering 30 & \centering 10 
    &	$32.52	\pm	0.57$	\\
    CFG-4& \centering 40 & \centering 30 & \centering 20 
    &	$10.44	\pm	0.32$	\\
    CFG-5& \centering 40 & \centering 25 & \centering 20 
    &	$12.36	\pm	0.35$	\\
\hline
  \end{tabular}
\end{table}

\begin{figure}[h]
\centering
\includegraphics[width=0.5\textwidth]{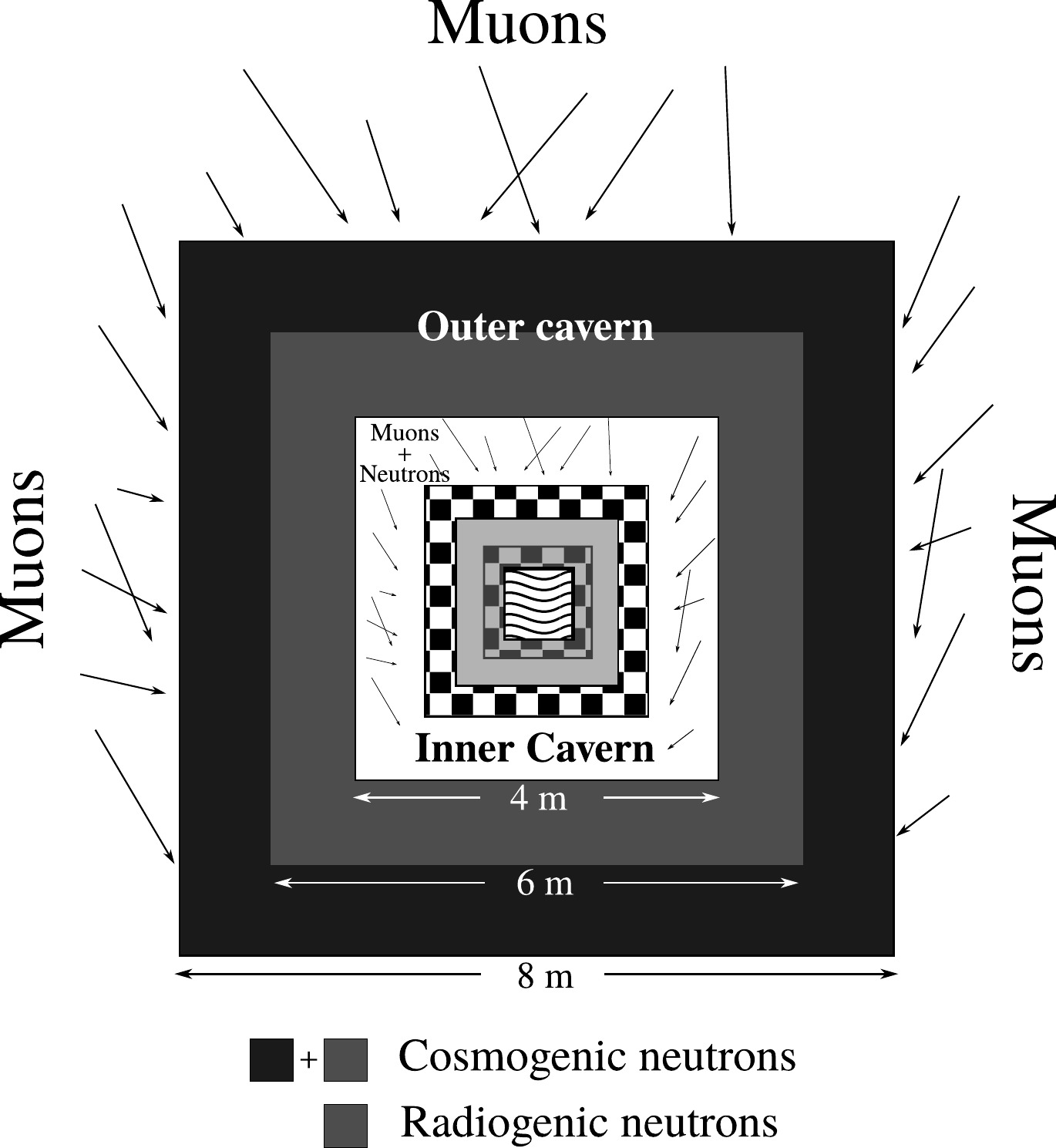}
\caption{{Schematic diagram of geometry used for simulation. The crystal, teflon and copper layers are shown together with wave pattern, PP2 is the black and light gray checkered region, Pb is the light grey region and PP1 is the black and white checkered region. The dark grey rectangular region is responsible for the radiogenic neutron background and black rectangular region is responsible for the cosmogenic neutron flux.}}
\label{fig:detful}
\end{figure}

About 47\% of the input neutrons are stopped by the 400 mm thick polypropylene. It is observed that a large number of neutrons are produced in Pb from interactions initiated by incoming neutrons. Neutron back-scattering is also large at the Pb boundary. A second polypropylene layer (PP2) is needed to attenuate the neutrons produced in Pb. From Table \ref{tab:shielddes}, we see that shielding combination CFG-4 provides the best neutron reduction. Neutrons, gamma, muons and electrons are the major backgrounds in a typical experimental setup. Muons and neutrons can also interact with the detector / shielding materials to produce more neutrons. We investigate the effect of shielding and calculate neutron flux at a detector using a simple geometry. The shielding design is based on the dimensions of shielding materials obtained from CFG-4 shown in Table \ref{tab:shielddes}. The geometry of the rectangular rock element and the experimental setup is given in Figure~\ref{fig:detful}. The experimental setup consists of a cylindrical CsI crystal with a radius of 2.2 cm and height 4 cm. Surrounding the crystal there are cylindrical layers of covering and shielding materials with various thicknesses: Teflon (0.05 cm), copper (0.6 cm), polypropylene (20 cm, PP1), Pb (30 cm), polypropylene (40 mm, PP1). There is rock block with thickness of 200 cm surrounding this experimental set up with $\sim100$ cm of air gap between them.

Muons can interact with the shielding materials to produce high energy neutrons having energies up to few GeVs. Shielding these neutrons are difficult. The knowledge of the muon flux is therefore very important for shielding design. The muon flux reaching the outer boundary of each layer of the experimental setup obtained from this simulation for 10 days of data are shown in the Table~\ref{tab:table4}. During passage of cosmic muons through the shield materials, almost no reduction of muon flux can be seen.
\begin{table}[h]
  \centering
  \caption{\label{tab:table4}The flux of muons at the top surface of different layers.}
 \begin{tabular}{c|c}
 \hline
 Material 
 & Flux $(\mathrm{cm^{-2}\thinspace s^{-1}}$)  \\ \hline
 PP1 
 & 4.45($\pm$0.24) $\times$ 10$^{-7}$ \\ \hline
 Pb 
 & 4.52($\pm$0.26) $\times$ 10$^{-7}$ \\ \hline
 PP2
 & 4.67($\pm$0.30) $\times$ 10$^{-7}$ \\ \hline
 Cu 
 & 3.82($\pm$1.12) $\times$ 10$^{-7}$ \\ \hline
 \end{tabular}
 \end{table}
 
The muons and muon induced neutrons are tracked through the successive layers of shielding. Neutrons produced in each layer of shielding is recorded. Neutrons are produced from both muon initiated interactions and neutron interactions. The neutron production rate in each layer is shown in Table~\ref{tab:table2}.  
\begin{table}[h]
  \centering
  \caption{\label{tab:table2}The production rate of neutrons in different layers. Uncertainties shown are statistical only.}
 \begin{tabular}{c|c}
 \hline
 Material 
 &  Neutron production rate ($\mathrm{cm^{-3}\thinspace s^{-1}}$)  \\ \hline
 PP1 
 & 2.05($\pm$0.04) $\times$ 10$^{-10}$ \\ \hline
 Pb 
 & 1.72($\pm$0.01) $\times$ 10$^{-8}$ \\ \hline
 PP2
 & 6.64($\pm$0.06) $\times$ 10$^{-10}$ \\ \hline
 \end{tabular}
 \end{table}
 
Only a fraction of neutrons produced in each layer reach the inner layer of detector. Others either get absorbed or scatter off. The neutrons which get reflected back from a layer (i) can be absorbed by the previous layer (ii) get transmitted out of the setup or (iii) get reflected back again to the same layer. 
These effects are taken into consideration to avoid multiple counting. Neutrons reaching each layer of the detector include neutrons produced in 
all previous layers. For instance neutrons reaching the copper layer include neutrons produced in rock, polypropylene and lead.

\begin{figure}[h]
\centering
\includegraphics[width=0.6\textwidth]{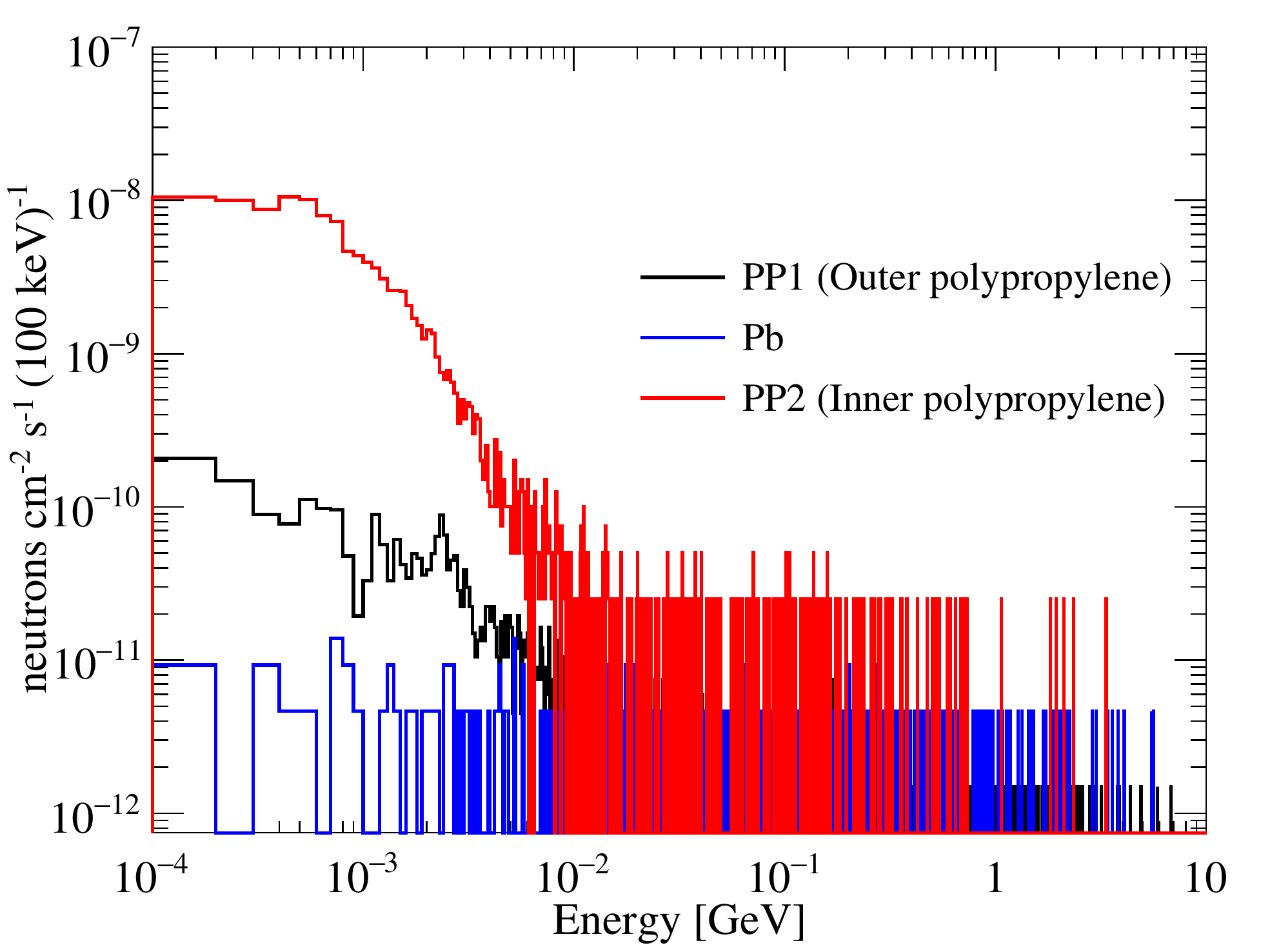}
\caption{Comparison of neutron energy distribution at different layers of experimental setup}
\label{fig:munelayer}
\end{figure}

The flux of cosmogenic neutrons estimated at the top surface of different layers for 20 days of data are shown in Figure ~\ref{fig:munelayer}. 
The flux of neutrons that reach the top surface of each layer of the experimental setup for 20 days of data is shown in Table~\ref{tab:table3}.
\begin{table}[h]
  \centering
   \caption{\label{tab:table3}The flux of neutrons at the top surface of each layer. Uncertainties shown are statistical only.}
 \begin{tabular}{c|c|c}
 \hline
 Material & $E_\mathrm{neutron}^\mathrm{mean}$ (MeV) & Flux ($\mathrm{cm^{-2}\thinspace s^{-1}}$) \\ \hline
 PP1 & 81 & 8.19($\pm$0.11) $\times$ 10$^{-9}$\\ \hline
 Pb & 280 & 3.04($\pm$0.12) $\times$ 10$^{-9}$\\ \hline
 PP2 & 8 & 1.44($\pm$0.02) $\times$ 10$^{-7}$\\ \hline
 Cu & 19 & 7.44($\pm$3.72) $\times$ 10$^{-8}$\\ \hline
 CsI & 9 & 6.15($\pm$4.35) $\times$ 10$^{-8}$\\ \hline
 \end{tabular}
 \end{table}
 
The increase in neutron flux at the boundary of PP2 is due to the production of new neutrons in the Pb layer. The increase in mean energy of neutrons at the Cu layer can be due to the absorption of lower energy neutrons by PP2. The mean scattering length of neutrons is smaller in hydrogen compared to other materials like C, Pb or Fe for neutron energies less than $\sim$10 MeV. Whereas for higher neutron energies, the mean scattering length increases compared to other materials \cite{CBungau}. Hence, the higher energy neutrons cannot be moderated easily using hydrogen-based shielding material.

\section{WIMP sensitivity estimate based on neutron background\label{sec:sens}}
We make estimates for sensitivity by following the method suggested in Ref. \cite{lewin}, and used in KIMS \cite{kimsfirst} DM search. Considering a dark matter halo model with a Maxwellian velocity distribution as described in Ref. \cite{lewin}, the total WIMP event-rate in recoil energy range between $E_{R_1}$ and $E_{R_2}$ is given by \cite{kimsfirst}
\begin{equation}\begin{split}
    R(v_E,v_\mathrm{esc}) &= \frac{k_0}{k_1}\int_{E_{R_1}}^{E_{R_2}} dE_R\left\{c_1\frac{R_0}{E_0r}e^{-c_2E_R/E_0r}-\frac{R_0}{E_0r}e^{-v^2_\mathrm{esc}/v_0^2}\right\},\\
    R_0&=5.47\left(\frac{\mathrm{GeV}/c^2}{m_\chi}\right)\left(\frac{\mathrm{GeV}/c^2}{m_t}\right)\left(\frac{\sigma_0}{\mathrm{pb}}\right)\left(\frac{\rho_\chi}{\mathrm{GeV}/c^2/\mathrm{cm}^3}\right)\left(\frac{v_0}{\mathrm{km/s}}\right),\\
    E_0&=\frac{1}{2}m_\chi v_0^2,\quad r=\frac{4m_\chi m_t}{(m_\chi+m_t)^2},    
    \end{split}
\end{equation}
where $m_\chi$ is the dark matter mass, $m_t$ is the mass of a target nucleus, $\rho_\chi = 0.3$ GeV$\thinspace$cm$^{-3}$ is the local dark matter density, $v_0 = 220$ km$\thinspace$s$^{-1}$ is a Maxwell velocity parameter, $v_\mathrm{esc} = 650$ km$\thinspace$s$^{-1}$ is the local galactic escape velocity of WIMP, $k_0/k_1\approx 1$ and $c_1 , c_2$ are constants which depend on the Earth (target) velocity $v_E$, relative to the dark matter distribution as discussed in Ref. \cite{lewin}. $\sigma_0$ is the WIMP-nucleus `zero momentum transfer' cross-section, and $R_0$ is the total event rate (in kg$^{-1}$day$^{-1}$) for $v_E$ = 0,  and $v_\mathrm{esc}=\infty$. Radiogenic neutrons were completely stopped by the shielding. Only cosmogenic neutrons were able to reach the crystal. From simulation, the total number of nuclear recoil events due to neutrons within the energy range 8-60 keV in the CsI crystal is estimated to be $\sim$6 kg$^{-1}$year$^{-1}$ (corresponding to a 90\% poisson CL of 12 kg$^{-1}$year$^{-1}$). The nuclear recoil energy scale could be converted into the electron equivalent energy scale using the quenching factors for CsI crystals reported in Ref. \cite{park}. This turns out to be $\sim$1.5 to 6.5 keV.

\begin{figure}[h]
\centering
\includegraphics[width=0.6\textwidth]{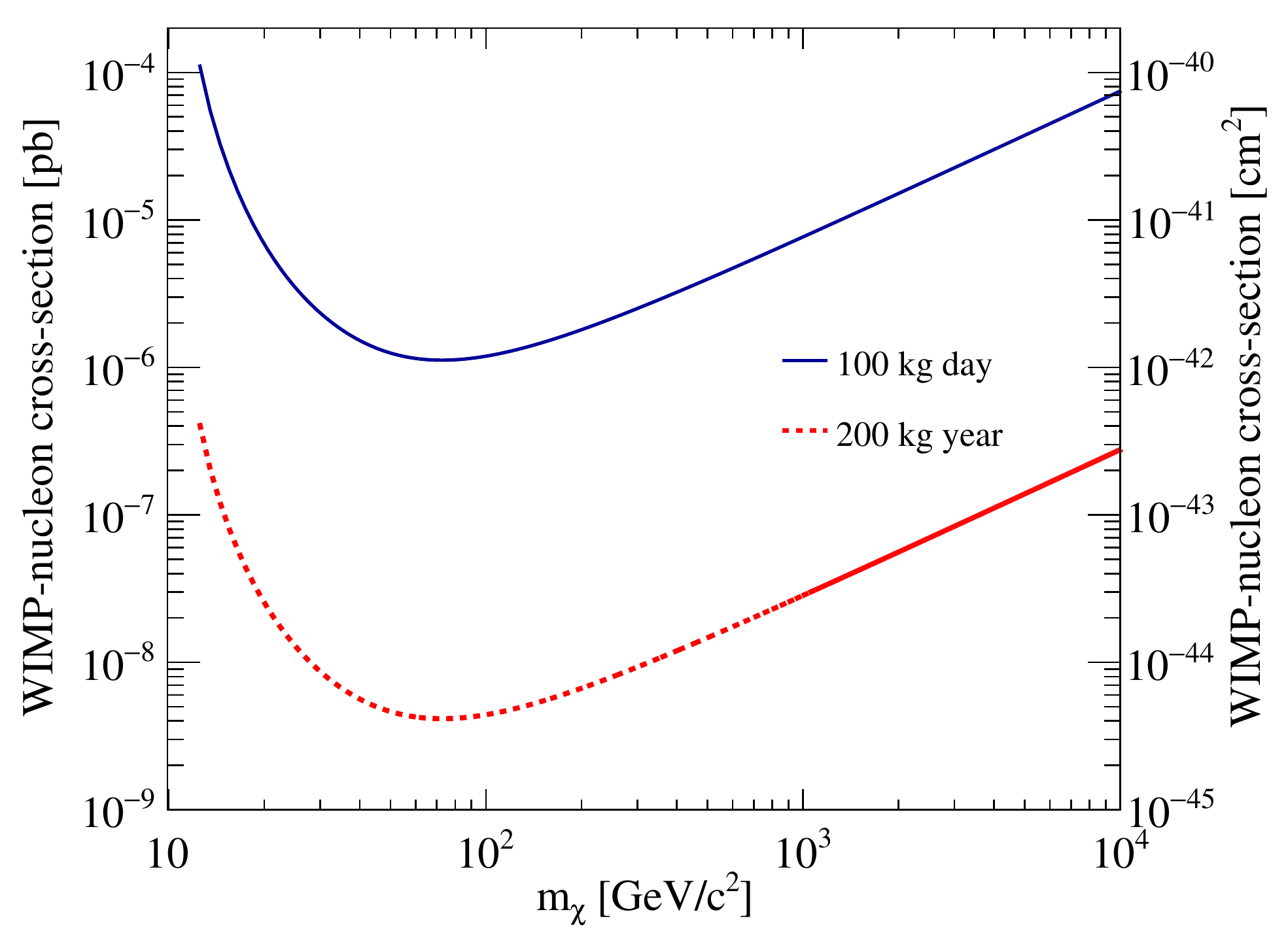}
\caption{Sensitivity of a CsI based detector at JUSL. Blue line shows the sensitivity for a 100 kg$\thinspace$day exposure (272 g detector for 1 year) and the red line shows the sensitivity for 200 kg$\thinspace$year exposure (200 kg detector for 1 year).}
\label{fig:sensitive}
\end{figure}

The WIMP-nucleon cross-section can be obtained from the WIMP-nucleus cross-section using the formula \cite{kimsfirst}
\begin{equation}
  \sigma_{W-n}=\sigma_{W-A}\frac{\mu_n^2}{\mu_A^2}\frac{C_n}{C_A}
\end{equation}
where $\sigma_{W-n}$ is the WIMP-nucleon cross-section, $\sigma_{W-A}$ is the WIMP-nucleus cross-section for nucleus of mass number $A$, $\mu_n$ and $\mu_A$ are the reduced masses of WIMP-nucleon and WIMP-nucleus systems respectively and $C_A/C_n=A^2$ for spin independent interaction \cite{kimsfirst}. Since the CsI crystal has two different nuclei, the combined limit on the WIMP-nucleon cross-section ($\sigma$) is given by \cite{kimsfirst}
\begin{equation}
  \frac{1}{\sigma}=\frac{1}{\sigma_\mathrm{Cs}}+\frac{1}{\sigma_\mathrm{I}},
\end{equation}
where $\sigma_\mathrm{Cs}$ and $\sigma_\mathrm{I}$ are the WIMP-nucleon cross sections limits for Cs and I nuclei respectively. Neglecting the background contribution from gamma and $\alpha$ particles, the estimated sensitivity of a CsI based direct dark matter search experiment at JUSL is shown in Figure \ref{fig:sensitive}. Spin independent interaction has been considered and the WIMP-nucleon cross-section is shown as a function of WIMP-mass. The blue line corresponds to a 272 g detector as considered in the simulation, running for 1 year. The red dotted line corresponds to a 200 kg detector running for 1 year assuming the same level of background. Sensitivity estimates in the present study only assume background events from neutrons; this leads to an optimistic estimate of the sensitivity. A more realistic calculation will require consideration of other backgrounds, which are not calculated in this study, along with neutrons. The parameters such as the quenching factors etc. have been assumed to be similar to that of KIMS. A better estimate of the sensitivity will be obtained if all the detector parameters are measured and understood specifically for DINO. Nevertheless, setting up a direct dark matter search experiment at JUSL is feasible. 

\section{Conclusion}
\label{sec:conc}
An estimation of the neutron flux due to radiogenic and cosmogenic sources for a dark matter search experiment at JUSL, an underground laboratory in India with 555 m rock overburden, has been done using GEANT4 framework. The radiogenic neutrons have energies upto few MeVs whereas muon-induced neutron energy extends up to few 10s of GeVs. A study has also been performed to find the optimal shielding combination for effective reduction of these neutron backgrounds.   

The specific neutron activity due to ($\alpha,n$) reactions from the surrounding rock materials has been obtained as $6.77\pm1.12\thinspace$yr$^{-1}\thinspace$g$^{-1}$ of rock from $^{238}$U and $5.33\pm0.90\thinspace$yr$^{-1}\thinspace$g$^{-1}$ of rock from $^{232}$Th. The specific neutron activity due to spontaneous fission of $^{238}$U is obtained as $3.43\pm0.55\thinspace$yr$^{-1}\thinspace$g$^{-1}$. The flux of radiogenic neutrons reaching the inner cavern is obtained as $1.12(\pm0.11)\times10^{-5}\thinspace$cm$^{-2}\thinspace$s$^{-1}$ above 100 keV energy threshold with a mean energy of 1.34 MeV and $5.75(\pm0.58)\times10^{-6}\thinspace$cm$^{-2}\thinspace$s$^{-1}$ above 1 MeV energy threshold with a mean energy of 2.18 MeV.

Cosmic muon events are generated on the surface using Gaisser's parametrization and are made to propagate through the rock material. It is observed that only about 7\% of the generated cosmogenic neutrons pass through the rock of thickness 2 m. The muon flux at the outer cavern is found to be nearly $4.49(\pm0.25)\times10^{-7}$ cm$^{-2}\thinspace$s$^{-1}$ with an average energy of about 200 GeV. These muons are propagated through rock to obtain the muon and muon induced neutron flux at the cavern. The muon flux at the boundary of the inner cavern is obtained to be $4.45(\pm 0.24)\times 10^{-7}$ cm$^{-2}\thinspace$s$^{-1}$. Muon induced neutron flux from rock in the inner cavern is found to be $0.93(\pm0.08)\times10^{-8}\thinspace$cm$^{-2}$s$^{-1}$ without any energy threshold and $7.25(\pm0.65)\times10^{-9}\thinspace$cm$^{-2}\thinspace$ s$^{-1}$ above 1 MeV. Our estimated values of neutron and muon fluxes are comparable with calculations for dark matter experiments in Boulby mine \cite{boulby}. The measured value of muon flux at the WIPP salt mine which is at a similar depth ($\sim1580$ m.w.e) is $4.77\times10^{-7}$ cm$^{-2}\thinspace$s$^{-1}$ \cite{WIPP}. Our estimation of muon induced neutron flux is comparable with their calculation (1.6 $\times$ 10$^{-8}\thinspace$cm$^{-2}\thinspace$s$^{-1}$) reported in the same paper.

The total neutron flux reaching the Inner cavern/laboratory from both radiogenic and muon induced reactions in rock is found to be 5.76($\pm0.58)\times$ 10$^{-6}\thinspace$cm$\thinspace^{-2}\thinspace$s$^{-1}$ mostly dominated by radiogenic neutrons. Neutrons produced from muon and neutron interaction with the shielding materials also contribute to the neutron flux at the detector. A high number of neutrons are produced in Pb.

Radiogenic neutrons are easily stopped by the 40 cm thick polypropylene layer. Cosmogenic neutrons can penetrate the shielding and reach the detector. Moreover, muons generate neutrons while traversing through the shielding. Muon and neutron fluxes have been estimated at various layers of shielding. It is seen that a shielding configuration comprising of Polypropylene (40 cm) + Pb (30 cm) + Polypropylene (20 cm) from the outside towards the experimental setup gives the best reduction in neutron background. 
 Since the neutron production rate is high in Pb, a second layer of polypropylene is needed for effective shielding of these neutrons. With the best determined shielding configuration, the flux of muon induced neutron at the detector is found to be 6.15($\pm$4.35) $\times$ 10$^{-8}$ cm$^{-2}\thinspace$s$^{-1}$.

The radiogenic neutron yield from contamination in the detector materials has not been considered in this calculation. D.M. Mei et al. \cite{Mei} showed that there is no $(\alpha,n)$ neutron yield in lead due to a very high Coulomb barrier.

The sensitivity of a CsI based WIMP dark matter search experiment at the Jaduguda mine has been estimated. The results show that a direct WIMP dark matter search experiment is feasible at JUSL.

\section*{Acknowledgements}
We would like to thank the Department of Atomic Energy (DAE), Government of India, for support through the project Research in Basic Sciences (Dark Matter). BM acknowledges the support of JC Bose National fellowship of the Department of Science and Technology (DST), Government of India. PB and SS acknowledge the support under Raja Ramanna fellowship of the Department of Atomic Energy (DAE), Government of India.
\bibliography{references}{}
\bibliographystyle{JHEP}

\end{document}